\documentclass[letterpaper]{aa}

\usepackage{graphicx} 
\usepackage{natbib}
\usepackage{txfonts}
\begin{document}
\title{VLT-ISAAC 3-5 $\mu$m spectroscopy of embedded young low-mass stars. III. Intermediate-mass sources in Vela\thanks{Based on observations collected at the European Southern Observatory at La Silla and Paranal, Chile (ESO Programme I64.I-0605)}}
\titlerunning{VLT-ISAAC spectroscopy of YSOs in Vela}
\author{W.-F.~Thi \inst{1,2,3} \and E.~F.~van~Dishoeck\inst{2} \and E.~Dartois\inst{4} \and K.~M.~Pontoppidan\inst{2} \and W.~A.~Schutte\inst{2} \and P.~Ehrenfreund\inst{2} \and L.~d'Hendecourt\inst{4} \and H.~J. Fraser\inst{2,5}}

\institute{Sterrenkundig Instituut Anton Pannekoek, University of
  Amsterdam, Kruislaan 403 1098 SJ Amsterdam, The Netherlands \and
  Leiden Observatory, P. O.  Box 9513, 2300, Leiden, The Netherlands
  \and ESA Research Fellow, Research Support Science Department,
  ESTEC, Keplerlaan 1, P. O.  2201 AZ, Noordwijk, The Netherlands \and
  Astrochimie Exp\'erimentale, Institut d'Astrophysique Spatiale,
  Universit\'e Paris-Sud, b\^at. 121, F-91405 Orsay, France \and
  Department of Physics, University of Strathclyde, John Anderson
  Building, 107 Rottenrow, Glasgow G4 ONG, Scotland}

\offprints{ewine@strw.leidenuniv.nl}

\date{Received ... ;Accepted 28/11/2005}

\abstract
    {}
   {We study in this paper the ice composition in the envelope around intermediate-mass class I Young Stellar Objects (YSOs).}
   {We performed a spectroscopic survey toward
    five intermediate-mass class I YSOs located in the Southern Vela molecular cloud in the $L$
    (2.85--4.0 $\mu$m) and $M$ (4.55--4.8 $\mu$m) bands at resolving
    powers $\lambda/\Delta \lambda=$~600-800 up to 10,000, using the
    Infrared Spectrometer and Array Camera mounted on the {\em Very
      Large Telescope-ANTU}. Lower mass companion objects were observed
    simultaneously in both bands.}
    {Solid
    H$_{\mathrm{2}}$O at 3~$\mu$m is detected in all sources,
    including the companion objects. CO ice at 4.67 $\mu$m is detected
    in a few main targets and one companion object. One object
    (\object{LLN 19}) shows little CO ice but strong 
     gas-phase CO ro-vibrational lines in absorption.  The CO ice profiles are
    different from source to source. The amount of water ice and CO
    ice trapped in a water-rich mantle may correlate with the flux
    ratio at 12 and 25 $\mu$m. The abundance of H$_2$O-rich CO likely
    correlates with that of water ice. A weak
    feature at 3.54 $\mu$m attributed to solid CH$_\mathrm{3}$OH and a
    broad feature near 4.62 $\mu$m are observed toward \object{LLN
    17}, but not toward the other sources. 
    The derived abundances
    of solid CH$_{\mathrm{3}}$OH and OCN$^{-}$ are $\sim$10 $\pm$ 2\%
    and $\sim$1 $\pm$ 0.2\% of the H$_{\mathrm{2}}$O ice abundance
    respectively. 
    The H$_{\mathrm{2}}$O optical depths do not show an increase with
    envelope mass, nor do they show lower values for the companion objects compared
    with the main protostar. The line-of-sight CO ice abundance does not
    correlate with the source bolometric luminosity.}
    {Comparison of the solid CO profile toward
    \object{LLN 17}, which shows an
    extremely broad CO ice feature, and that of its lower mass
    companion at a few thousand AU, which exhibits a narrow profile,
    together with the detection of OCN$^{-}$ toward \object{LLN 17}
    provide direct evidences for local thermal processing of the ice.}

\maketitle
%
\section{Introduction}

Dust grains play an important role in the evolution of clouds from
protostellar cores to circumstellar disks. Since dust grains are the
main source of opacity, they control the thermal balance of clouds.
The surfaces of cold grains act as heat sink for highly exothermic
reactions to occur (e.g., formation of H$_2$) or provide sites for
atoms and molecules to freeze on.  The freeze-out of molecules like CO
is found to be important for regulating the gas phase chemistry of
other species (e.g., \citealt{Bergin1997ApJ...486..316B};
\citealt{Bergin2001ApJ...557..209B};
\citealt{Jorgensen2004A&A...416..603J}). The frozen atoms and
molecules accumulate on top of a refractory core (silicates,
carbonaceous compounds) and form an icy mantle or react with other
species to synthesize more complex molecules.  The relative chemical
composition of this ice mantle is well determined after three decades
of studies using both ground-based and space-borne telescopes.  The
core-mantle grain model is supported by spectropolarimetry studies
(e.g., \citealt{Holloway2002MNRAS.336..425H}). Solid
H$_{\mathrm{2}}$O, CO, and CO$_{\mathrm{2}}$ abound in most lines of
sight where ices are detected \citep{deGraauw1996A&A...315L.345D}.
Sometimes, minor species such as CH$_{\mathrm{4}}$ ($\sim$2\%), HCOOH
($\sim$2\%), OCN$^{-}$ ($\sim$0.2--1\%) (e.g.,
\citealt{vanBroekhuizen2005A&A}), and H$_{\mathrm 2}$CO (3--6\%) are
found (e.g., \citealt{Boogert1998A&A...336..352B} ;
\citealt{Keane2001A&A...376..254K};
\citealt{Boogert2004ApJS..154..359B}).  By contrast, the presence of
other minor constituents such as NH$_{\mathrm{3}}$ is controversial;
its abundance relative to water ice is likely less than $\sim$10\%
(e.g., \citealt{Ehrenfreund2000ARA&A..38..427E};
\citealt{Dartois2001A&A...365..144D};
\citealt{Dartois2002A&A...394.1057D};
\citealt{Lacy1998ApJ...501L.105L};
\citealt{Taban2003A&A...399..169T}). It should be emphasized that most
of these studies refer to high mass young stellar objects.

Solid methanol (CH$_{\mathrm{3}}$OH) is a particular case. It
epitomizes the importance of molecular solids in the understanding of
the gas phase chemistry. The radiative association of CH$_3^+$ and
H$_2$O is an inefficient gas-phase process that yields methanol
abundances of $\sim~10^{-9}$ relative to H$_2$ while abundances of
10$^{-7}$--10$^{-6}$ have been found in hot cores (e.g.,
\citealt{Blake1987ApJ...315..621B};
\citealt{Sutton1995ApJS...97..455S}). The prevalent view is that the
high abundance of gas phase methanol come from the release of large
amounts of frozen methanol formed on grain surfaces
(\citealt{Charnley1995ApJ...448..232C};
\citealt{vanderTak2000A&A...361..327V};
\citealt{Horn2004ApJ...611..605H}).  Methanol ice abundance relative
to water ice is found to vary from less than 3\% w.r.t. water ice in
quiescent regions up to 30\% around massive protostars
\citep{Dartois1999A&A...342L..32D}. If all the methanol ice is
released in the gas phase, the abundance of methanol in the gas phase
with respect to H$_2$ will amount to 3 $\times$ 10$^{-7}$ -- 3
$\times$ 10$^{-6}$ assuming that the abundance of water ice with
respect to H$_2$ is $\sim$10$^{-5}$ (e.g.,
\citealt{Whittet2003dge..conf.....W}). A similar situation is found
for low-mass objects.  \cite{Chiar1996ApJ...472..665C} set a stringent
limit of 5\% of methanol with respect to water ice for sources located
in the Taurus molecular cloud, while
\cite{Pontoppidan2003A&A...404L..17P} found abundant methanol ice
(14--25\% of water ice) in 4 out of $\simeq$ 40 envelopes around
protostars observed with the VLT. Possible formation routes of solid
methanol are also disputed. The formation rate of solid methanol by
hydrogenation of CO ice in the absence of energy input (i.e. hot
atoms, UV or particle irradiation) measured in laboratory experiments
remains controversial (e.g., \citealt{Hiraoka2002ApJ...577..265H};
\citealt{Hiraoka2005ApJ...620..542H};
\citealt{Watanabe2004ApJ...616..638W}).

The advent of 8m class telescopes equipped with large format arrays
opens up the opportunity to study large samples of low and
intermediate-mass sources. We present here the first observations of
molecular ice features in the $L$ (2.8--4.1 $\mu$m) and $M$ (4.5--5.1
$\mu$m) bands toward class I intermediate-mass young stellar objects
(YSOs) located in the Vela molecular cloud complex.  The spectra were
obtained in the context of a large programme using the {\it Infrared
  Spectrometer And Array Camera} (ISAAC) mounted at the {\it Very
  Large Telescope} ANTU (VLT-ANTU) of the {\it European Southern
  Observatory} (ESO). Two major absorption features are observable
with ground-based telescopes, along with some weak features.  The
first strong feature is centered around 3.01 $\mu$m ($\sim$ 3300
cm$^{-1}$) and is usually attributed to the stretching mode of solid
H$_{\rm 2}$O.  The study of the solid-water profile has been used to
better explore the ice structure in the water matrix (e.g.,
\citealt{Smith1989ApJ...344..413S};
\citealt{Smith1993MNRAS.263..749S};
\citealt{Maldoni1998MNRAS.298..251M}). The feature shows a broad
excess absorption beyond 3.2 $\mu$m whose origin is still unclear,
although scattering by the larger grains (0.1--1 $\mu$m in radius) in
the size distribution is the best candidate
(\citealt{Smith1989ApJ...344..413S};
\citealt{Dartois2001A&A...365..144D}).

The other important feature is the solid-CO band at 4.67 $\mu$m (2140
cm$^{-1}$), whose profile is sensitive to the shape and size of the
grains as well as the ice composition and temperature and is
therefore a diagnostic of the evolutionary state and thermal history
of ices \citep{Sandford1988ApJ...329..498S}. As soon as the grain is
warmed to $\simeq$~12--15~K by the luminosity of the object, CO
molecules can diffuse into the ice and form new bonds, changing the
morphology of the mixture and thus the profile at 4.67 $\mu$m, or they
can sublime back to the gas phase
(\citealt{Al-Halabi2004A&A...422..777A};
\citealt{Collings2003ApJ...583.1058C}; \citealt{Givan1997JPhysChem}).
The mobility of CO and its high abundance in cold icy mantles also
explain why it is a key species for surface reactions leading to
polyatomic molecules such as CO$_{\rm 2}$ and CH$_{\rm 3}$OH
(\citealt{Rodgers2003ApJ...585..355R};
\citealt{Chiar1998ApJ...498..716C};
\citealt{Teixeira1998A&A...330..711T}).

The VLT-ISAAC observations are used to constrain the physical and
chemical conditions in the envelopes of a sample of intermediate-mass
stars.  The paper is organized as follows. We present the objects of
our sample in Sect.~\ref{vela:obj} and the observations and data
reduction procedures in Sect.~\ref{vela:obs}.  The results on the
water ice band are described in Sect.~\ref{vela:L_band}. Evidence for
solid CH$_{\rm 3}$OH is presented in Sect.~\ref{vela:methanol}.  The
CO-ice band is presented in Sect.~\ref{vela:co}. A possible
correlation between the abundance of CO embedded into a water ice
matrix and the IRAS 12~$\mu$m/ 25~$\mu$m color ratio is discussed in
Sect.~\ref{vela:cloud}. Conclusion are provided in
Sect.~\ref{vela:conclusion}. These data complement the survey of CO
and other species for a sample of $\sim$40 low mass YSO's by
\cite{Pontoppidan2003A&A...404L..17P,Pontoppidan2003A&A...408..981P}
obtained in the same programme.

\begin{figure}[ht]
\centering
  \resizebox{\hsize}{!}{\includegraphics[]{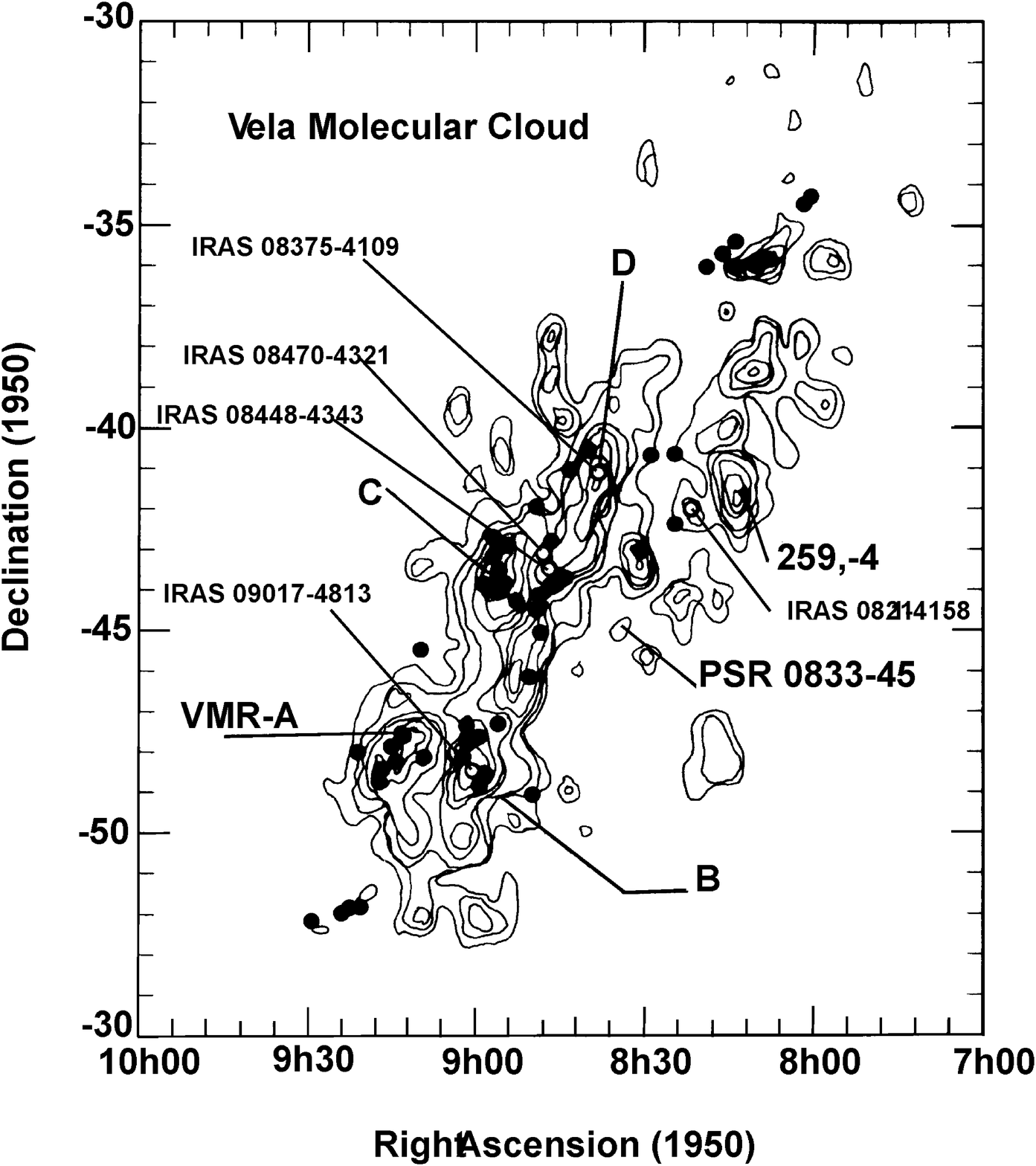}}
\caption[Position of the observed sources]{CO (1--0) contour map of the Vela Molecular Cloud complex \citep{Murphy1991A&A...247..202M}. The complex is composed of 4 clouds named A, B, C and D in the map.
  The positions of the observed sources are indicated and the filled
  circles are the location of the 50 other embedded young stellar
  objects.  Indicated as well are the locations of the Vela Pulsar and
  the H{\sc II} region 259,-4. The map is an adaptation of Fig.1 of
  \cite{Liseau1992A&A...265..577L}.}
\label{vela:fig_vela_cloud}
\end{figure}    

\section{Protostars in the Vela molecular clouds}\label{vela:obj}

We chose to observe 5 intermediate-mass protostars with bolometric
luminosities ${L_{*}\sim}$ 100--1000 ${L_\odot}$, corresponding to
masses of 2--10 $M_{\odot}$ \citep{Palla1993ApJ...418..414P}.  The
characteristics of the objects are summarized in
Table~\ref{vela:table_objects}. These objects are the brightest
sources in the $L$- and $M$-band taken from the sample of Class I YSO
candidates in the Vela molecular cloud
\citep{Liseau1992A&A...265..577L}, hereafter LLN. The location of the
sources is shown in Fig.~\ref{vela:fig_vela_cloud}, together with that
of other candidates not studied in this work. Four of these sources
have been observed by \cite{Pontoppidan2003A&A...408..981P}. The Vela
molecular cloud is located in the Galactic plane, and is reachable
only from observatories in the Southern hemisphere. It lies in an RA
range ($\sim$9h) which was not covered by the {\it Infrared Space
    Observatory} ({\it ISO}) satellite during the entire mission to
  avoid damage from the infrared radiation from the Earth.

The complex hosts three supernova remnants (\object{Vela SNR},
\object{Puppis A} and \object{G266.2 -1.2}), the latter discovered by
recent X-ray satellites \citep{Aschenbach1998Natur.396..141A}.  The
spectral energy distributions (SEDs) of the objects, red IRAS colors
and a near-infrared slope $s \equiv d\log(\lambda
F_{\lambda})/d\log(\lambda) \geq 0$ as defined by
\cite{Adams1987ApJ...312..788A} are characteristic of YSOs.
Near-infrared imaging has revealed that the chosen targets are the
brightest members of their respective proto-stellar clusters
\citep{Massi1999A&AS..136..471M}.  The companion stars observed
serendipitously (see Sect.~\ref{vela:obs}) are fainter class I
objects.  The entire complex was mapped in the CO
$J=\!1\!\rightarrow\!0$ transition by
\cite{Murphy1991A&A...247..202M}.  The complex is composed of at least
four Giant Molecular Clouds (A, B, C and D), with individual masses
exceeding 10$^5$ $M_\odot$, where active star formation occurs.  These
dark clouds are relatively nearby ($700 \pm 200$~pc), therefore giving
the possibility to study the ices in a typical galactic environment.
The location of the different objects and of their companions
coincides with a maximum in the CO $J=\!1\!\rightarrow\!0$ integrated
emission and with the highest visual extinction. Extinction studies
toward background stars have shown little foreground absorption from
diffuse clouds with a maximum of $A_V$=3.4
\citep{Reed1990AJ....100..156R}. Assuming that the threshold
extinction for detection of the water ice feature in this cloud is
$\sim$3.3, the amount of water ice that resides in the foreground is
negligible. The exact threshold value is unknown for Vela, but it
could be much higher than 3.3 \citep{Whittet2003dge..conf.....W}.

 \begin{table*}
\centering
\caption[Objects characteristics]{Characteristics of the sources.\label{vela:table_objects}}
\begin{tabular}[!ht]{llllllllll} 
\hline 
\hline
\noalign{\smallskip} 
& & \multicolumn{1}{c}{(1)}&\multicolumn{1}{c}{(2)}&\multicolumn{1}{c}{(3)}&\multicolumn{1}{c}{(4)}&\multicolumn{1}{c}{(5)}&\multicolumn{1}{c}{(6)}&\multicolumn{1}{c}{(7)}&\multicolumn{1}{c}{(8)}\\
\multicolumn{2}{c}{Source}&\multicolumn{1}{c}{$L_{\mathrm{NIR}}$}&\multicolumn{1}{c}{$L_{\mathrm{MIR}}$}&\multicolumn{1}{c}{$L_{\mathrm{IRAS}}$}&\multicolumn{1}{c}{$L_{\mathrm{submm}}$}
& \multicolumn{1}{c}{$L_{bol}$} &\multicolumn{1}{c}{$M_{\mathrm{env}}$}&\multicolumn{1}{c}{$A_{\mathrm{V}}$}& \multicolumn{1}{c}{$N_*$}\\
&\multicolumn{1}{c}{}&\multicolumn{1}{c}{(\%)} &\multicolumn{1}{c}{(\%)} &\multicolumn{1}{c}{(\%)} &\multicolumn{1}{c}{(\%)} &\multicolumn{1}{c}{($L_{\odot}$)} &\multicolumn{1}{c}{($M_{\odot}$)}&\multicolumn{1}{c}{(Mag)}&\multicolumn{1}{c}{(pc$^{-2}$)}\\
\noalign{\smallskip} 
\hline
\noalign{\smallskip} 
\object{LLN 8}& \object{IRAS 08211--4158}   &  5.3 &  10.5 & 53.3 & 31.0  & 141\  $^{(a)}$ & \multicolumn{1}{c}{...}    &  \multicolumn{1}{c}{...}   &  \multicolumn{1}{c}{...} \\
\object{LLN 13}& \object{IRAS 08375--4109}  &  5.3 & \ 6.8 & 72.8 & 17.7  & 960\  $^{(b)}$ & 2.6$^{(b)}$ & $\geq$~40$^{(a)}$ &800$^{(b)}$  \\
\object{LLN 17}& \object{IRAS 08448--4343}  &  1.3 & \ 4.0 & 60.8 & 33.9  & 3100\ $^{(b)}$ & 6.4$^{(b)}$ & $\sim$~40$^{(a)}$&3400$^{(b)}$ \\
\object{LLN 19}& \object{IRAS 08470--4321}  &  4.8 &  20.3 & 65.8 & \ 9.1 & 1600\ $^{(b)}$ & 3.5$^{(b)}$ & $\sim$~40$^{(a)}$&2400$^{(b)}$ \\
\object{LLN 41}& \object{IRAS 09017--4716}  &  ... &  ...  & 80.7 & 19.3      &  470              & ...    & ...   \\
\noalign{\smallskip} 
\hline
\noalign{\smallskip} 
\object{LLN 20}$^{(c)}$& \object{IRAS 08476--4306}   &  0.5     & 3.1      &  66.8    &  29.7  &  1600\ $^{(b)}$ & 2.2$^{(b)}$  &  &  3900$^{(b)}$\\
\object{LLN 33}$^{(c)}$& \object{IRAS 08576--4314}   &  2.6     & 12.6      & 80.6     & 4.2 & 91$^{(a)}$   & ...  & ...  & ... \\
\object{LLN 39}$^{(c)}$& \object{IRAS 09014--4736}   &  1.3      & 5.7      & 71.1     & 22.0  & 807$^{(a)}$  & ...  & ...  & ... \\
\object{LLN 47}$^{(c)}$& \object{IRAS 09094--4522}   &  3.4     & 13.3      & 74.4     & 8.9   &  21$^{(a)}$  & ...  & ...  & ... \\
\noalign{\smallskip} 
\hline
\end{tabular}
\ \\
\begin{flushleft}
  {\em Notes:} The dots refer to values which are not available in the literature.\\
  
  {\em Column 1--4:} 
  The fractional luminosities are computed by
  \cite{Liseau1992A&A...265..577L}. The indices refer to the
  wavelength range 1-5 $\mu$m (column 1), 5-12 $\mu$m (column 2),
  12-100 $\mu$m (column 3) and 0.1-1 mm (column 4) for NIR, MIR, IRAS
  and submm, respectively. $L_{\mathrm{submm}}$ are estimated by
  \cite{Liseau1992A&A...265..577L} when the millimeter observations do
  not exist.\\
  
  {\em Column 5:} Bolometric luminosity.
  
  {\em Column 6:} The envelope masses ($M_{\mathrm{env}}$) are derived
  from 1 mm continuum flux by \cite{Massi1999A&AS..136..471M} for four
  of our sources. Envelope masses are only given for those sources for
  which millimeter observations have been carried out.\\
    
  {\em Column 7:} Extinction estimated by \citet{Liseau1992A&A...265..577L}.\\
  
  {\em Column 8:} $N_*$ refers to the maximum stellar surface
  densities measured.\\
  
  {\em Reference:} $^a$\cite{Liseau1992A&A...265..577L},
  $^b$\cite{Massi1999A&AS..136..471M}, $^c$ {\em VLT-ISAAC} data
  analyzed by \cite{Pontoppidan2003A&A...408..981P}.
\end{flushleft}
\end{table*}

\section{Observations and data reduction}\label{vela:obs}

\begin{figure}[ht] 
\centering
  \resizebox{\hsize}{!}{\includegraphics{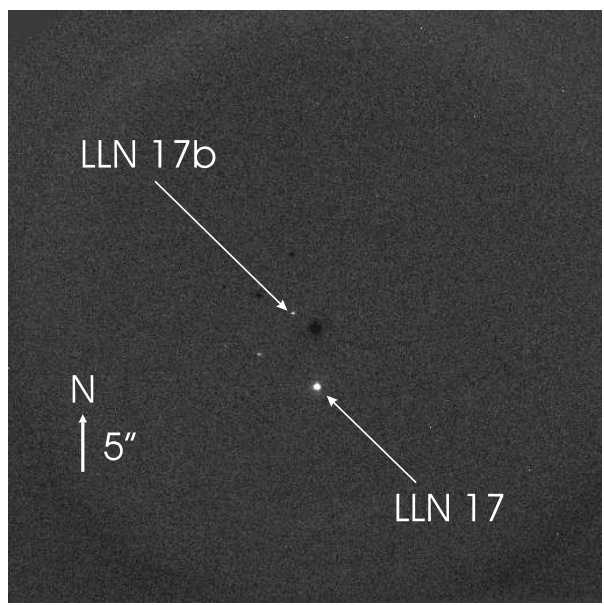}}
\caption[]{Acquisition image in the stellar field of \object{LLN 17} (\object{IRAS 08448--4343}. The orientation of the image (1024 $\times$ 1024 pixels) is North-South and the pixel scale is 1 \arcsec /pixel and corresponds to the length of
    the arrow. The negative points are the result of the chopping.
  The main target is called the source IRS 17-57 in
  \cite{Massi1999A&AS..136..471M} and the companion object is source
  IRS 17-40. \label{vela:acqui}}
\end{figure}

The observations were performed using ISAAC mounted on VLT-ANTU in
Paranal, Chile, in the $L$ (2.85--4.1~$\mu$m) and $M$ (4.5-4.8~$\mu$m)
band in January and November 2001 under mediocre conditions.  ISAAC is
a cryogenic mid-infrared (1--5 $\mu$m) imaging facility and grating
spectrometer.  The instrument uses a Santa Barbara 1024 $\times$ 1024
pixel InSb array. The entrance slit of the grating spectrometer was
set to 0.6\arcsec, which results in a resolving power of 600 in the
$L$ band and 800 in the $M$-band (low resolution mode). Since the
average seeing at Paranal in the infrared was $\sim$0.6$''$, little
flux from the object was lost. The slit was oriented such that the
main target and a second nearby object, when present, could be
observed simultaneously.  An example of an acquisition image at
$L$-band is shown in Fig.~\ref{vela:acqui}. The companion object is
located between 5 and 10\arcsec\ from the main source. The large
detector combined with the 0.6\arcsec\ slit allowed the entire $L$-
and $M$- band to be obtained in one exposure each. The integration
times of 30--40 min per object and by band were chosen such that the
$S/N$ ratio reaches 60 in the wavelength regions of good atmospheric
transmission, allowing an analysis of the profile.  Any other
absorption features with an optical depth of 0.05 (3$\sigma$) can be
detected as well.  In complement, three sources \object{LLN 13}
(\object{IRAS 08375--4109}), \object{LLN 17} (\object{IRAS
  08448--4343}) and \object{LLN 19} (\object{IRAS 08470--4321}) were
observed at a resolving power of 10,000 (0.3\arcsec\ slit) in the
$M$-band. A summary of the observations is presented in
Table~\ref{vela:table_log}.

The objects were observed with a chopping along the slit with a throw
of typically 15\arcsec\ together with nodding. This technique removes
the majority of the sky and background noise. After or before each
science target, the spectrum of a standard star was obtained with the
same setting (\object{BS~3185} of spectral type F6II and
\object{BS~3842} of spectral type G8II).  The difference in airmass
between the target and standard star is always kept less than 0.015.
The stars used for atmospheric correction were chosen to be late-type
photometric standards. The targets have accurate published photometric
measurements (see Table~\ref{vela:table_log}) and those values were
used to estimate the flux of the main and companion objects. The
companions are 3.5 to 18 times fainter than the main sources, which
are the brightest objects in their respective proto-clusters. The
absolute flux calibration is uncertain by 10\%.  The standard star
spectra were not modeled and therefore hydrogen absorption lines in
the spectrum were not removed.  In consequence, apparent emission
features at 3.04 $\mu$m (H{\sc I} Pf$\epsilon$ 5-10), 3.30 $\mu$m
(H{\sc I} Pf$\delta$ 5-9), and 3.74 $\mu$m (H{\sc I} Pf$\gamma$ 5-8)
in the data may be spurious features from the rationing by the
standard star.

\begin{figure*}[ht] 
\centering
  \resizebox{\hsize}{!}{\includegraphics{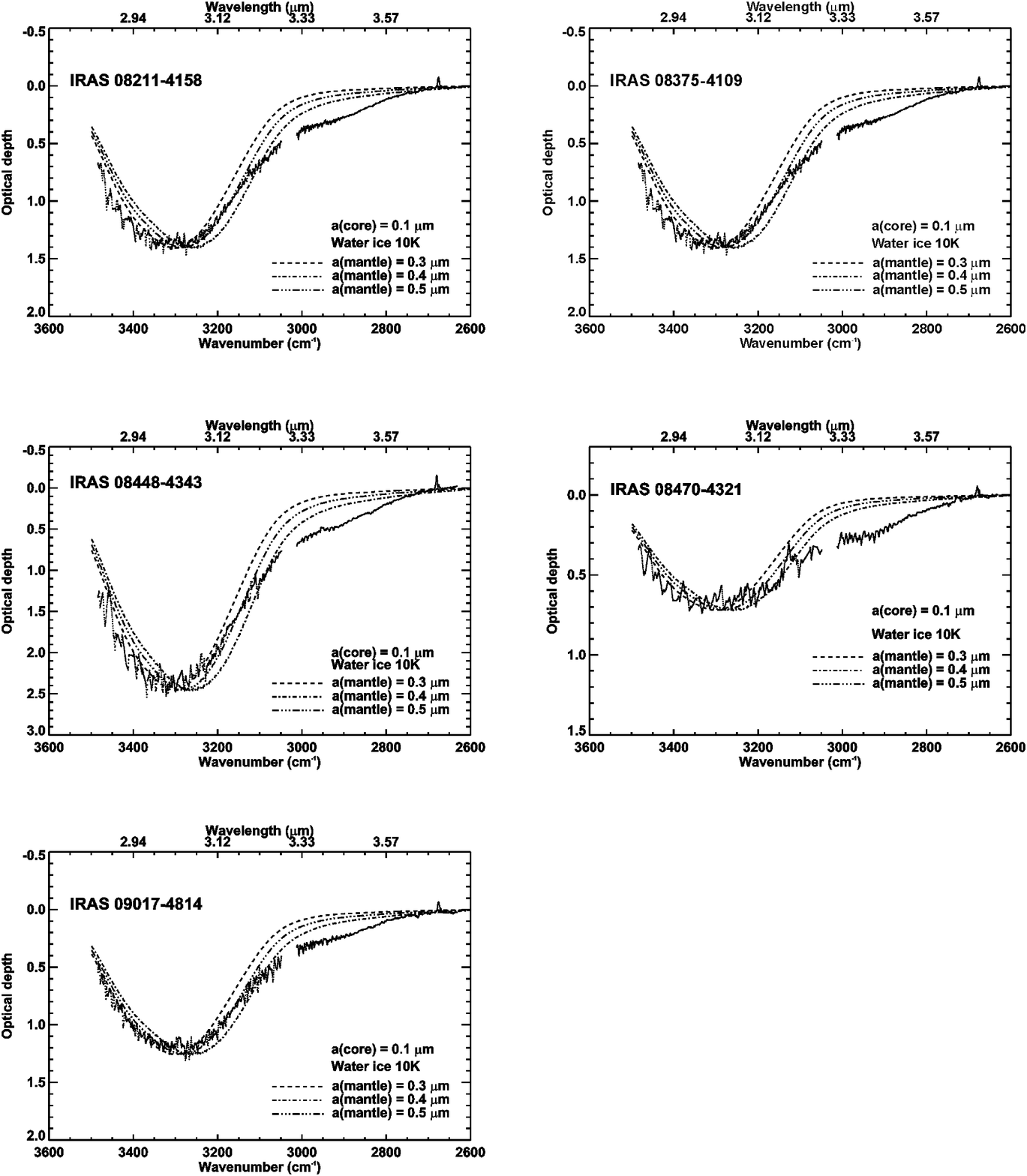}}
\caption[Optical depth profiles of the 3 $\mu$m water ice feature for the
main objects]{Optical depth profiles of the 3 $\mu$m ice feature for
  the 5 objects (dotted line) compared to ice coated grains at
  different radii using laboratory spectra taken from
  \citep{Hudgins1993ApJS...86..713H}.}
\label{vela:fig_h2o1}
\end{figure*}

The spectra were reduced using in-house software written in the
Interactive Data Language ({\em IDL}).  The data reduction procedure
is standard for IR observations (bad pixel maps, flat-fielding, etc.).
The distortions of the raw spectra were corrected.  In the $L$-band,
the wavelength calibration was obtained from arc lamp measurements.
In the $M$-band, because of the scarcity of halogen lamp lines in this
wavelength range, the spectra were calibrated by comparing the strong
atmospheric absorption lines with high signal-to-noise ratio spectra
of the atmosphere above Paranal measured by the ESO staff. The data
points in the strongest telluric absorption features show the lowest
signal-to-noise ratio and were therefore dropped. This is particularly
the case for the telluric gaseous methane absorption band around 3.25
$\mu$m. Error bars are omitted in the flux plots for clarity for the
$L$-band spectra, but the atmospheric transmission is provided for
information in Fig.~\ref{vela:fig_h2o1}.

\begin{table*}[ht]
\centering
\caption[Summary of the observations]{Summary of the observations and infrared fluxes taken from the literature (The companion objects are located within 30\arcsec\ from the main source).
\label{vela:table_log}}
\begin{tabular}[!ht]{llllllllll}
\hline
\hline
\noalign{\smallskip} 
&\multicolumn{2}{c}{Coordinate}&\multicolumn{1}{c}{$J$}&\multicolumn{1}{c}{$H$}&\multicolumn{1}{c}{$K$}&\multicolumn{1}{c}{$L'$}&\multicolumn{1}{c}{$M$} &

\multicolumn{1}{c}{Standard}&\multicolumn{1}{c}{Resolution$^b$} \\
\multicolumn{1}{c}{Source}&\multicolumn{1}{c}{RA (J1950)}&\multicolumn{1}{c}{Dec (J1950)}&\multicolumn{1}{c}{(Mag)$^a$}&\multicolumn{1}{c}{(Mag)$^a$}&\multicolumn{1}{c}{(Mag)$^a$}&\multicolumn{1}{c}{(Mag)}&\multicolumn{1}{c}{(Mag)$^a$}&\multicolumn{1}{c}{star}\\
\noalign{\smallskip} 
\hline
\noalign{\smallskip} 
\object{LLN 8}   & 08 21 07.4 & $-$41 58 13  & \phantom{$>$}13.00 & 10.12 & 7.59 &4.79    & 4.01  &  \object{BS~3185}& Low\\
\object{LLN 8b}  & 08 21 08.1 & $-$41 57 44   & ... & ... & ...  & ...     & ...  &  \object{BS~3185}& Low\\
\object{LLN 13}  & 08 37 30.8 & $-$41 09 29   & $>$14.80 & 12.14 & 8.95 & 5.87    & 5.09 &  \object{BS~3842}& Low, Med.\\
\object{LLN 13b} & ...        & ...           & ... & ...& ...   & ...     & ...  &  \object{BS~3842}& Low\\
\object{LLN 17} & 08 44 48.8 & $-$43 43 26   & \phantom{$>$}14.0 & 11.60 & 9.13 &  5.77     & 4.7  &  \object{BS~3185}& Low, Med.\\
\object{LLN 17b} & ...      & ... & ...& ...  & ...     & ...     & ...  &  \object{BS~3185} & Low\\
\object{LLN 19}   & 08 47 01.4 & $-$43 21 25   & \phantom{$>$}14.7 & 11.92 & 8.88 & 4.08     & 2.73  &  \object{BS~3185}& Low, Med.\\
\object{LLN 41} & 09 01 44.0 & $-$47 17 19   & $>$14.80 & 11.51 & 8.15 & 4.5     & 3.3  &  \object{BS~3185}& Low \\
\object{LLN 41b}  & 09 01 42.3 & $-$47 16 43   & \phantom{$>$}14.1 & 11.31 & 9.41 & 7.3     & 6.9  &  \object{BS~3185}& Low\\
\noalign{\smallskip} 
\hline
\noalign{\smallskip} 
\object{LLN 20}$^c$  & 08 47 39.4 & $-$43 06 08 & $>$14.90 & 12.73 & 10.80 & 7.7 & 6.2 & ...& Med.\\
\object{LLN 33}$^c$  & 08 57 36.8 & $-$43 14 35 & $>$16.20 & 14.1 & 11.16 & 7.19 & 5.7 & ...& Med.\\
\object{LLN 39}$^c$  & 09 01 26.4 & $-$47 36 34 & \phantom{$>$}11.79 & 10.16 & 8.55 & 5.64 & 5.2 & ...& Med.\\
\object{LLN 47}$^c$  & 09 09 25.6 & $-$45 22 51 & \phantom{$>$}11.82 & 10.10 & 8.90 & 5.94 & 4.8 & ...& Med.\\
\noalign{\smallskip} 
\hline
\noalign{\smallskip} 
\end{tabular}
\ \\
\begin{flushleft}
{\em Notes:}\\ 
$^a$ Photometry taken from \cite{Liseau1992A&A...265..577L}.\\ 
$^b$ Low resolution $R\simeq$800 in the $L$ and $M$ band; Medium resolution $R\simeq$10,000 in the $M$ band only.\\
$^c$ Observed by \cite{Pontoppidan2003A&A...408..981P}.
\end{flushleft}
\end{table*}

\section{Observational results}

\subsection{Water ice}\label{vela:L_band}

\begin{figure*}[ht]
  \centering
  \resizebox{\hsize}{!}{\includegraphics{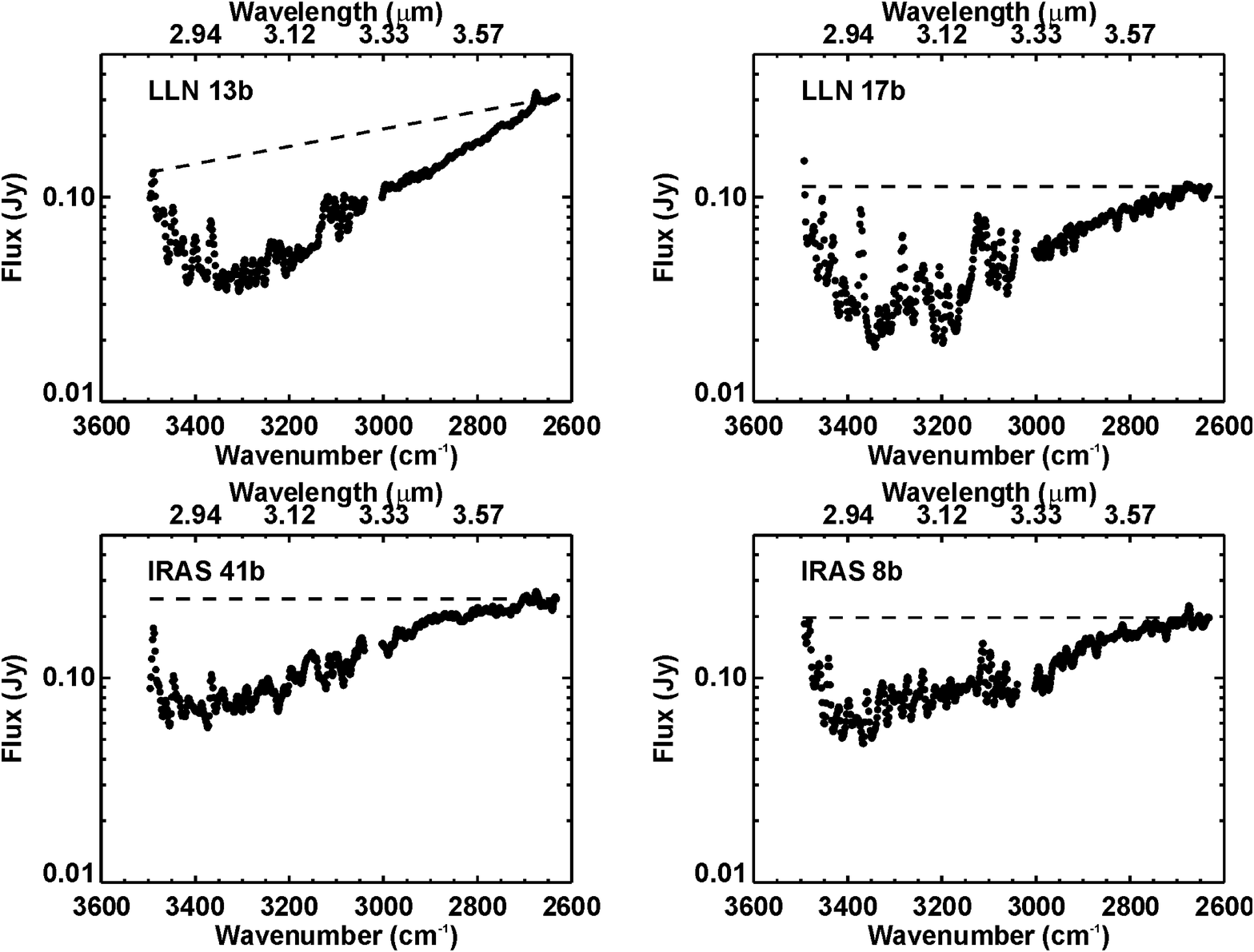}}
\caption{$L$-band spectra of the companion objects. The water ice absorption feature is seen in all spectra. The spectra are dominated by the spurious features due to a bad telluric transmission correction. The dashed lines indicate the adopted continuum used to estimate the water-ice optical depth.}
\label{vela:fig_companion_Lband}
\end{figure*}

The spectra of the main objects, converted into optical depth scale,
are presented in Fig.~\ref{vela:fig_h2o1}, while the $L$-band spectra
of the companion objects are displayed in
Fig.~\ref{vela:fig_companion_Lband} in flux scale.  A local blackbody
that fits the photometric data at 2.5 $\mu$m and 3.8 $\mu$m, measured
by \cite{Liseau1992A&A...265..577L} was subtracted from the $L$-band
spectra to obtain the optical depth scale. The low signal-to-noise
ratio in the spectra of the companion objects prevents detailed
analysis of the profiles, although column densities can be derived.

A broad absorption feature extending from 3600 to 2600 cm$^{-1}$ (2.85
to 3.8~$\mu$m) is observed in all spectra including the companion
objects. This feature, centered around 3.1~${\mathrm \mu}$m, is
attributed to the O--H stretching mode of ${\rm H_2O}$ ice (e.g.,
\citealt{Tielens1983JPhCh..87.4220T}).  The detection of methanol-ice
toward \object{LLN 17} (\object{IRAS 08448--4343}) at 3.54 $\mu$m is
discussed in Sect.~\ref{vela:methanol}.  The band shape is comparable
to that found in quiescent clouds, for example in the line of sight of
\object{Elias 16} \citep{Smith1989ApJ...344..413S}, a field star
located behind the Taurus cloud, which traces unprocessed ices.
Table~\ref{Vela:table_water_results} shows the optical depths of the
water band for the main and companion objects. The optical depths
toward the companion objects are grouped around 1.1 $\pm$ 0.3, whereas
a larger scatter is seen for the main objects.  In the remainder of
this section, we focus our analysis on the main objects.

\begin{figure}[ht] 
  \centering
  \resizebox{\hsize}{!}{\includegraphics{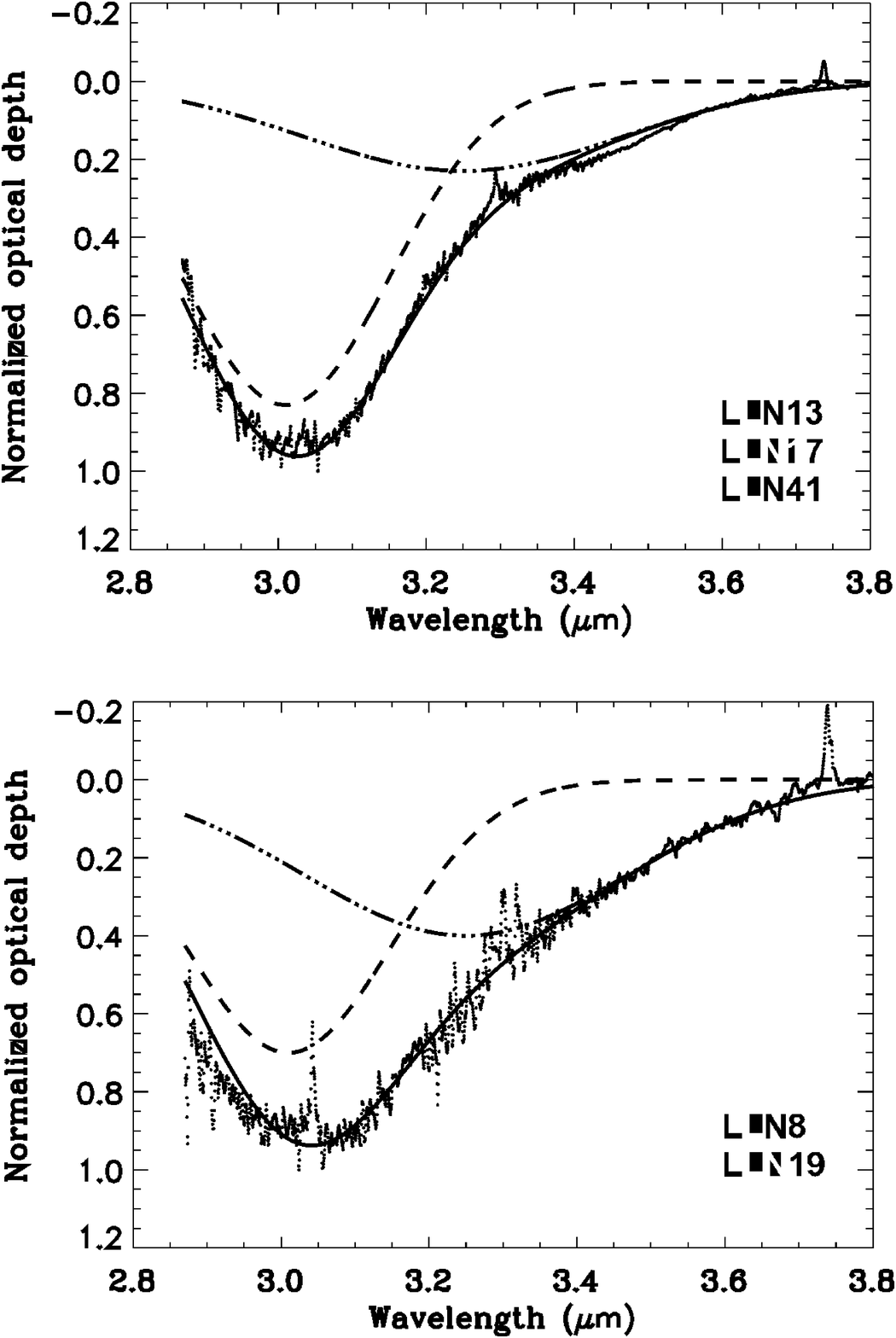}}
\caption[]{Normalized optical depths of the Vela sources fitted by a
  combination of 2 Gaussians, one centered at 3.01 $\mu$m, the other
  one at 3.25 $\mu$m. The upper panel shows the normalized optical
  depth for 3 Vela sources. The spectra are overplotted one on each
  other. The similarity between the 3 sources is so strong that it is
  difficult to separate the sources in the plot. The lower panel shows
  the spectrum of 2 other Vela sources.  The spectra in both panels
  are fitted by a linear combination of two Gaussians.
\label{vela:2families}}
\end{figure}

An extended wing to the main feature is present in all our sources.
The identification of the carrier which gives rise to this long
wavelength wing is a long-standing problem
(\citealt{Smith1989ApJ...344..413S};
\citealt{Dartois2001A&A...365..144D}). To compare the strength of the
wing relative to the main feature, the $L$-band spectra in optical
depth scale were normalized to unity at 3.01 $\mu$m. It appears that
the spectra can be divided into two groups. The first group is
composed of objects where the long-wavelength wing centered at 3.25
$\mu$m is relatively weak compared to the main feature.  Three objects
fall into this group (\object{LLN 13}, \object{LLN 17} and \object{LLN
  41}). Objects in the second group show a stronger wing relative to
the main feature. Two objects in our sample (\object{LLN 8} and
\object{LLN 19}) pertain to this group.  The similarity in the band
profile among objects within the same group is remarkable
(Fig.~\ref{vela:2families}).  In both groups, the entire profile was
fitted empirically with a linear combination of two Gaussians centered
respectively at 3.01 ($FWHM$=0.14 $\mu$m) and 3.25$\mu$m ($FWHM$=0.44
$\mu$m).

\begin{figure}[ht]
\centering
\resizebox{\hsize}{!}{\includegraphics{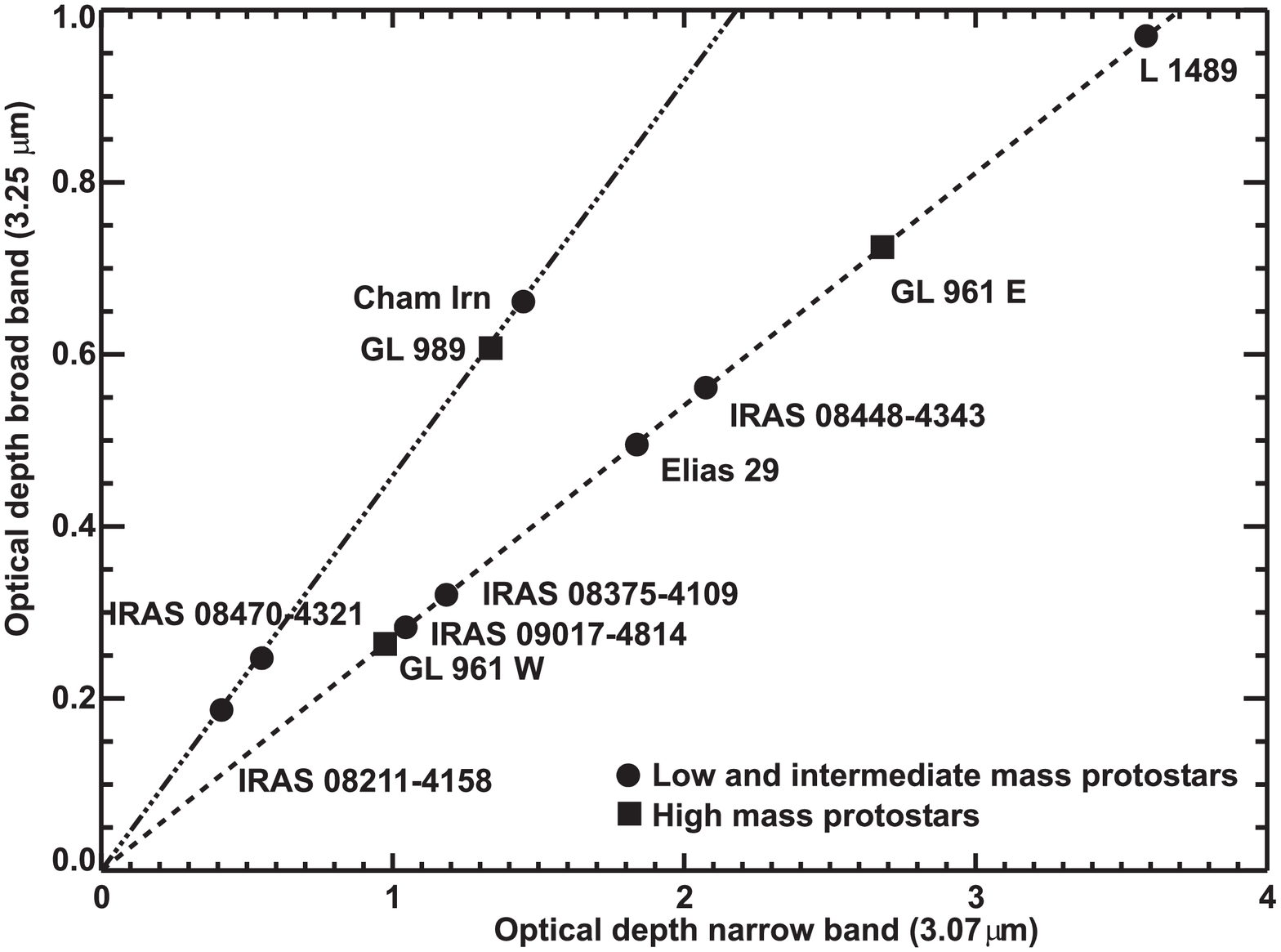}}
\caption{Optical depth at 3.25 $\mu$m versus optical depth at
  3.07$\mu$m. The straight lines can be fitted to the data. The origin
  of both lines is (0,0). The data for sources which are not located
  in the Vela cloud are taken from the ISO-archive.
\label{vela:fig_family}}
\end{figure}

A few objects exhibit an additional broad and shallow absorption
feature at 3.47 $\mu$m. The optical depth of this feature is reported
in Table~\ref{Vela:table_water_results}. The 3.47 $\mu$m feature is
also found in higher mass protostars \citep{Brooke1996ApJ...459..209B,
  Brooke1999ApJ...517..883B}. The broad 3.47 $\mu$m feature is
attributed to an hydrate (e.g.  ammonia hydrate,
\citealt{Dartois2002A&A...394.1057D}), in which other contributors
such as sp$^3$ carbon can be present
\citep{Allamandola1992ApJ...399..134A}. The optical depth of the main
feature at 3.01 $\mu$m is compared to that at 3.25 $\mu$m in
Fig.~\ref{vela:fig_family}. In addition, data of a selected number of
other low- and high-mass protostars are included. A strong correlation
can be found among objects in each group. The optical depth at 3.25
$\mu$m is found to be 0.27 and 0.45 times that at 3.01 $\mu$m for the
two groups, respectively. The tight correlations suggest that the
carrier(s) of the extended wing are strongly related to water-ice.
However, the correlations may be fortuitous since the number of
objects in the sample is limited.

\begin{table*}[ht]
\centering  
\caption[Water band fits with laboratory absorbance spectra]{Ice column densities. The errors are at 3~$\sigma$ level and are dominated by systematics. \label{Vela:table_water_results}}
\begin{tabular}[!ht]{lllllll}
\hline
\hline
\noalign{\smallskip} 
\multicolumn{2}{c}{Source} & \multicolumn{1}{c}{$\tau$(H$_2$O$_{\mathrm{ice}}$)}&\multicolumn{1}{c}{$N_{\mathrm{ice}}$(H$_2$O)}&\multicolumn{1}{c}{$T_{\mathrm{ice}}$(H$_2$O)}&\multicolumn{1}{c}{$N_{\mathrm{ice}}$(OCN$^{-}$)$^a$}&\multicolumn{1}{c}{$N_{\mathrm{ice}}$(CH$_3$OH)$^b$}\\
 & & &\multicolumn{1}{c}{(10$^{18}$ cm$^{-2}$)} &\multicolumn{1}{c}{(K)}&\multicolumn{1}{c}{(10$^{16}$ cm$^{-2}$)}&\multicolumn{1}{c}{(10$^{17}$ cm$^{-2}$)}\\
\noalign{\smallskip} 
\hline
\noalign{\smallskip} 
\object{LLN 8}  &\object{IRAS 08211--4158}  & 0.55 $\pm$ 0.10 & 0.80 $\pm$ 0.15 & 10--40 & $<$1.5& $<$1.0\\
\object{LLN 8b}  &\object{IRAS 08211--4158b} & 1.2\phantom{0} $\pm$ 0.3\phantom{0}  & 1.7\phantom{0} $\pm$ 0.4& 10--40 & ... & ... \\
\object{LLN 13}  &\object{IRAS 08375--4109}  & 1.4\phantom{0} $\pm$ 0.3\phantom{0} & 2.2\phantom{0} $\pm$ 0.4 & 10--40 & $<$0.5& $<$1.0 \\
\object{LLN 13b}  &\object{IRAS 08375--4109b} & 1.1\phantom{0} $\pm$ 0.3\phantom{0} &  1.7\phantom{0} $\pm$ 0.4& 10--40 & ... &...  \\
\object{LLN 17}  &\object{IRAS 08448--4343}  & 2.46 $\pm$ 0.50 & 3.61 $\pm$ 0.70 & 10--40 & \phantom{$<$}4.3 $\pm$ 0.5&\phantom{$<$}2.5 $\pm$ 1.0 \\
\object{LLN 17b}  &\object{IRAS 08448--4343b} & 1.4\phantom{0} $\pm$ 0.3 &  2.2\phantom{0} $\pm$ 0.2& 10--40 & $<$1.5& ...  \\
\object{LLN 19}  &\object{IRAS 08470--4321}  & 0.72 $\pm$ 0.10\phantom{0} & 1.05 $\pm$ 0.2 & 10--40 & $<$0.5& $<$1.0 \\
\object{LLN 41}  &\object{IRAS 09017-4716}  & 1.26 $\pm$ 0.20\phantom{0} & 1.85 $\pm$ 0.4 & 10--40 & $<$1.5&$<$1.0 \\ 
\object{LLN 41b}  &\object{IRAS 09017-4716b} & 0.8\phantom{0} $\pm$ 0.3\phantom{0} &  1.2\phantom{0} $\pm$ 0.4& 10--40 & $<$1.5& ...  \\  
\noalign{\smallskip} 
\hline
\end{tabular}
\ \\
\begin{flushleft}
{\em Notes:}\\ 
$^a$ The 3~$\sigma$ upper limits are derived assuming $\Delta \nu$=25 cm$^{-1}$ and $A$=1.0 $\times$ 10$^{-16}$ cm$^{-1}$ molec$^{-1}$. Upper limits derived from low resolution spectra are higher than that from medium resolution because of the possible contamination by CO gas phase lines.\\ 
$^b$ The 3~$\sigma$ upper limits are derived assuming $\Delta \nu$=14 cm$^{-1}$ and $A$=2.8 $\times$ 10$^{-18}$ cm$^{-1}$ molec$^{-1}$. No upper limits are given for the companion objects because the signal-to-noise ratios are too low to provide scientifically meaningful upper limits.\\
\end{flushleft}
\end{table*}

\begin{table*}[ht]
\centering  
\caption[]{Ice abundances relative to water ice. \label{Vela:table_ice_abundances}}
\begin{tabular}[!ht]{lllll}
\hline
\hline
\noalign{\smallskip} 
\multicolumn{1}{c}{Source} &\multicolumn{1}{c}{$N_{\mathrm{ice}}$(H$_2$O)}&\multicolumn{1}{c}{$N_{\mathrm{ice}}$(OCN$^{-}$)/}&\multicolumn{1}{c}{$N_{\mathrm{ice}}$(CH$_3$OH)/}&\multicolumn{1}{c}{$N_{\mathrm{ice}}$(CO)/} \\
                           &\multicolumn{1}{c}{(10$^{18}$ cm$^{-2}$)}&\multicolumn{1}{c}{$N_{\mathrm{ice}}$(H$_2$O)}&\multicolumn{1}{c}{$N_{\mathrm{ice}}$(H$_2$O)}&\multicolumn{1}{c}{$N_{\mathrm{ice}}$(H$_2$O)}\\
&&\multicolumn{1}{c}{(\%)}&\multicolumn{1}{c}{(\%)}&\multicolumn{1}{c}{(\%)}\\
\noalign{\smallskip} 
\hline
\noalign{\smallskip} 
\object{LLN 8}   & 0.80 $\pm$ 0.15          & $<$0.2 & $<$12.5&   $<$22\\
\object{LLN 8b}  & 1.7\phantom{0} $\pm$ 0.4 & ... & ... & ... \\
\object{LLN 13}  & 2.2\phantom{0} $\pm$ 0.4 & $<$0.2& $<$4.5 & \phantom{$<$}55 \\
\object{LLN 13b} & 1.7\phantom{0} $\pm$ 0.4 & ... & ... & ... \\
\object{LLN 17}  & 3.61 $\pm$ 0.70          & \phantom{$<$}1.7 & \phantom{$<$}6.9 & \phantom{$<$}15\\
\object{LLN 17b} & 2.2\phantom{0} $\pm$ 0.2 & $<$1.1 & ... & \phantom{$<$}17 \\
\object{LLN 19}  & 1.05 $\pm$ 0.2           & $<$0.7 & $<$9.5 & \phantom{$<$}4\\
\object{LLN 41}  & 1.85 $\pm$ 0.4           & $<$0.1 & $<$5.4 & \phantom{$<$}22\\ 
\object{LLN 41b} & 1.2\phantom{0} $\pm$ 0.4 & $<$1.9 & ... & $<$15 \\
\noalign{\smallskip} 
\hline
\end{tabular}
\end{table*}

It has long been realized that light scattering by large ice grains
leads to additional extinction on the long wavelength wing of the
water band (e.g., \citealt{Smith1989ApJ...344..413S}).  Several models
of water ice have been developed in which part of the long-wavelength
wing is attributed to scattering due to large grains (greater than
0.2~$\mu$m). Our model is similar to that previously used by
\cite{Smith1989ApJ...344..413S}. The silicate core radius is fixed at
a constant value of 0.1~$\mu$m.  The grain core is coated with a water
ice mantle whose thickness is allowed to vary to match the observed
spectra.  To simplify the problem, the ice is assumed to have a single
temperature.  The absorption and scattering cross sections are
computed using a Mie scattering theory program for coated-spheres
\citep{Bohren1983asls.book.....B}. The optical constants provided by
\citet{Hudgins1993ApJS...86..713H} for water ice and by
\cite{Draine2003ARA&A..41..241D} for the silicate core were used.  The
computed spectra for total grain radii of 0.3, 0.4 and 0.5 $\mu$m are
shown overlaid on the astronomical spectra in
Fig.~\ref{vela:fig_h2o1}.  The peak wavelength of the scattering cross
section is red-shifted compared to that of the absorption
cross-section and can therefore account for part of the
long-wavelength wing. The shape of the long-wavelength wing strongly
depends on the actual grain radius.  The maximum radius derived from
this model is 0.4--0.5 $\mu$m and the ice temperature is below 40~K.

The column densities were estimated by integrating numerically the
laboratory spectra over the water band from 2.8 to 3.8~$\mu$m using
the band strength for pure H$_2$O ice of $A$ = 2.0 $\times$ 10$^{-16}$
cm molec$^{-1}$ at 10~K measured by \cite{Hagen1981JChPh..75.4198H}:
\begin{equation}
N_{\rm solid}({\rm H_2O})=\int \frac{\tau_{\nu}\ d\nu}{A}
\end{equation}
where $\tau_{\nu}$ is the optical depth at wavenumber $\nu$
(cm$^{-1}$) and $A$ (cm molec$^{-1}$) is the integrated absorption
cross section per molecule (band strength).  The temperature and
column densities of the ice giving the best fits are summarized in
Table~\ref{Vela:table_water_results}. Uncertainties in experimental
band strengths for water ice are much lower than that in the
determination of the continuum around the water ice band.  The
inferred column densities varies from 0.8 to 3.6 $\times$ 10$^{18}$
cm$^{-2}$.


\subsection{Methanol ice}
\label{vela:methanol}

\begin{figure}[ht]
  \centering
  \resizebox{\hsize}{!}{\includegraphics{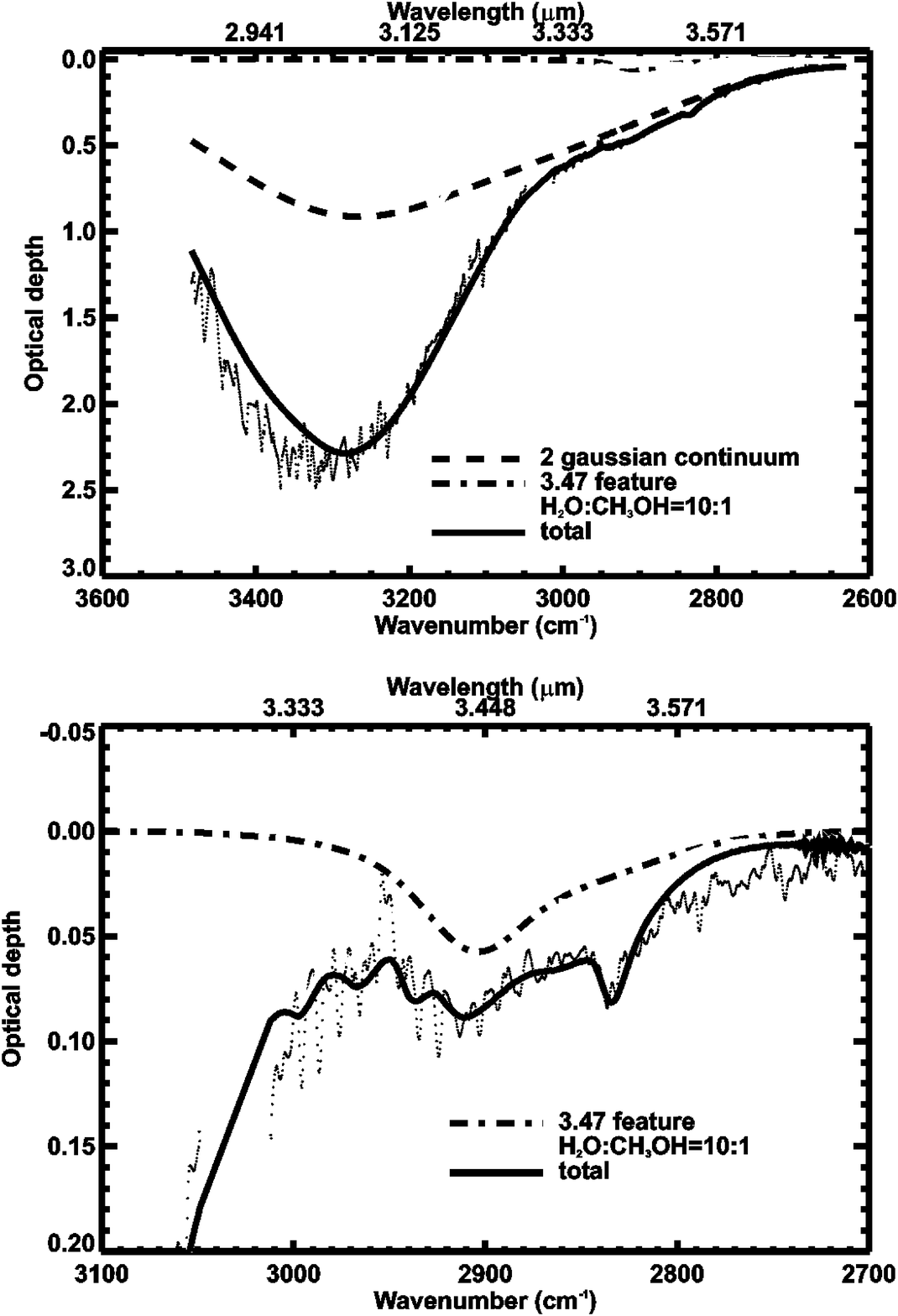}}
\caption{The upper panel displays the water ice feature toward \object{LLN 17}. The adopted 2 gaussian continuum, the 3.47 $\mu$m feature and the methanol ice:water ice = 1:10 mixture are overplotted. The lower panel shows a zoom around the methanol feature. In this plot, the 2 gaussian continuum has been subtracted.\label{vela:fig_methanol}}
\end{figure}

\begin{figure}[ht]
  \centering

  \resizebox{\hsize}{!}{\includegraphics{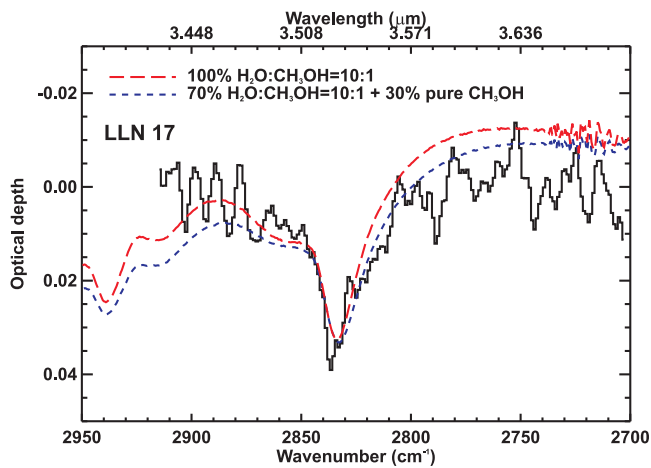}}
\caption{The methanol feature toward \object{LLN 17} is fitted
    by a single methanol mixture (CH$_3$OH:H$_2$O=1:10) and by a 2
    component mixture (70\% CH$_3$OH:H$_2$O=1:10 + 30\% pure
    CH$_3$OH). The improvement with the 2 component mixture is
    marginal.\label{vela:fig_methanol2_comparison}}
\end{figure}

Only \object{LLN 17} (\object{IRAS 08448--4343}) exhibits a strong
methanol feature at 3.54 $\mu$m superimposed on the solid water
absorption. The analysis of the methanol feature requires the
subtraction of the extended red absorption of the water ice and of the
3.47 $\mu$m. An artificial subtraction of the water ice wing is needed
because laboratory mixtures containing methanol ice do not take
scattering effects into account. The red-wing continuum is composed of
two gaussians whose characteristics are given in
Sect.~\ref{vela:L_band}.  The 3.47 $\mu$m feature is modeled
empirically by the sum of two gaussians as well ($\lambda_o$ = 3.47
$\mu$m, $FWHM$=0.12 $\mu$m and $\lambda_o$=3.44 $\mu$m, $FWHM$=0.48
$\mu$m).  The remaining feature is compared with mixed ices of H$_2$O
and CH$_{\mathrm{3}}$OH with different relative abundances.  The peak
position and the profile of the solid methanol bands are known to vary
with the amount of water in the mixture.  The presence of water shifts
the peak to higher frequency and narrows the methanol band.  The best
fit is obtained with the mixture H$_2$O:CH$_{\mathrm{3}}$OH=10:1. The
different components of the fit and the total are shown in the upper
limit of Fig.~\ref{vela:fig_methanol}. Possible improvement of the fit
with a two phase ice is tested. A two phase ice mantle composed of
pure methanol and a mixture H$_2$O:CH$_{\mathrm{3}}$OH=10:1 in the
proportion 30\% and 70\% respectively improves marginally the fit
(Fig.~\ref{vela:fig_methanol2_comparison}).

Adopting the integrated absorption coefficient valid for pure methanol
($A$= 2.8 $\times$ 10$^{18}$ cm molec$^{-1}$,
\citealt{Kerkhof1999A&A...346..990K}), the derived column density of
methanol is (2.5 $\pm$ 1.0) $\times$ 10$^{17}$ cm$^{-2}$ ($3\sigma$), 
corresponding to a
relative abundance compared to water ice of $\simeq$~6.9 $\pm$ 2\%.
The search toward the other objects was unsuccessful with an upper
limit of $\tau$(CH$_{\mathrm{3}}$OH)$<$0.02 ($3\sigma$),
corresponding to limits on the CH$_{\mathrm{3}}$OH ice abundances of
$\simeq$ 1 $\times$ 10$^{17}$ cm$^{-2}$ or a limit on the abundance of
$<$4.5--12.5\%. The methanol ice abundances and upper limits are
summarized in Table~\ref{Vela:table_ice_abundances}. The relatively
high upper limits (4.5--12.5\%) stem from the lower water ice
abundance in the line of sight of most objects ($N$(H$_2$O)~$<$~2.3
$\times$ 10$^{18}$ cm$^{-2}$) compared to that of \object{LLN 17}
($N$(H$_2$O)~=~3.6 $\times$ 10$^{18}$ cm$^{-2}$).

The detection of solid methanol toward the intermediate-mass protostar
\object{LLN 17} (\object{IRAS 08448--4343}) with a relative abundance
with respect to the water ice of $\simeq$~6.9\% and the non-detection
($<$~5--10\%) toward the other objects can be compared with the
variable abundances toward high-mass protostars (from $<$1~\% up to
$\sim$~30\% in \object{RAFGL 7009S} and \object{W33 A};
\citealt{Dartois1999A&A...342L..32D};
\citealt{Brooke1999ApJ...517..883B};
\citealt{Chiar1996ApJ...472..665C};
\citealt{Allamandola1992ApJ...399..134A}) as well as low mass
protostars ($<$5 up to 25\%,
\citealt{Pontoppidan2003A&A...404L..17P}). Some of the large
variations in the methanol ice abundance occur between objects of
similar mass located within the same cluster for low-mass star-forming
regions such as Serpens. The upper limits are relatively high
because of the low water ice column density found toward most
objects. Deeper limits for the LLN sources are needed 
to determine whether similar CH$_3$OH ice abundance
variations also hold for the Vela star-forming region.

\subsection{CO ice}
\label{vela:co}

\begin{figure*}
\resizebox{\hsize}{!}{\includegraphics[angle=90]{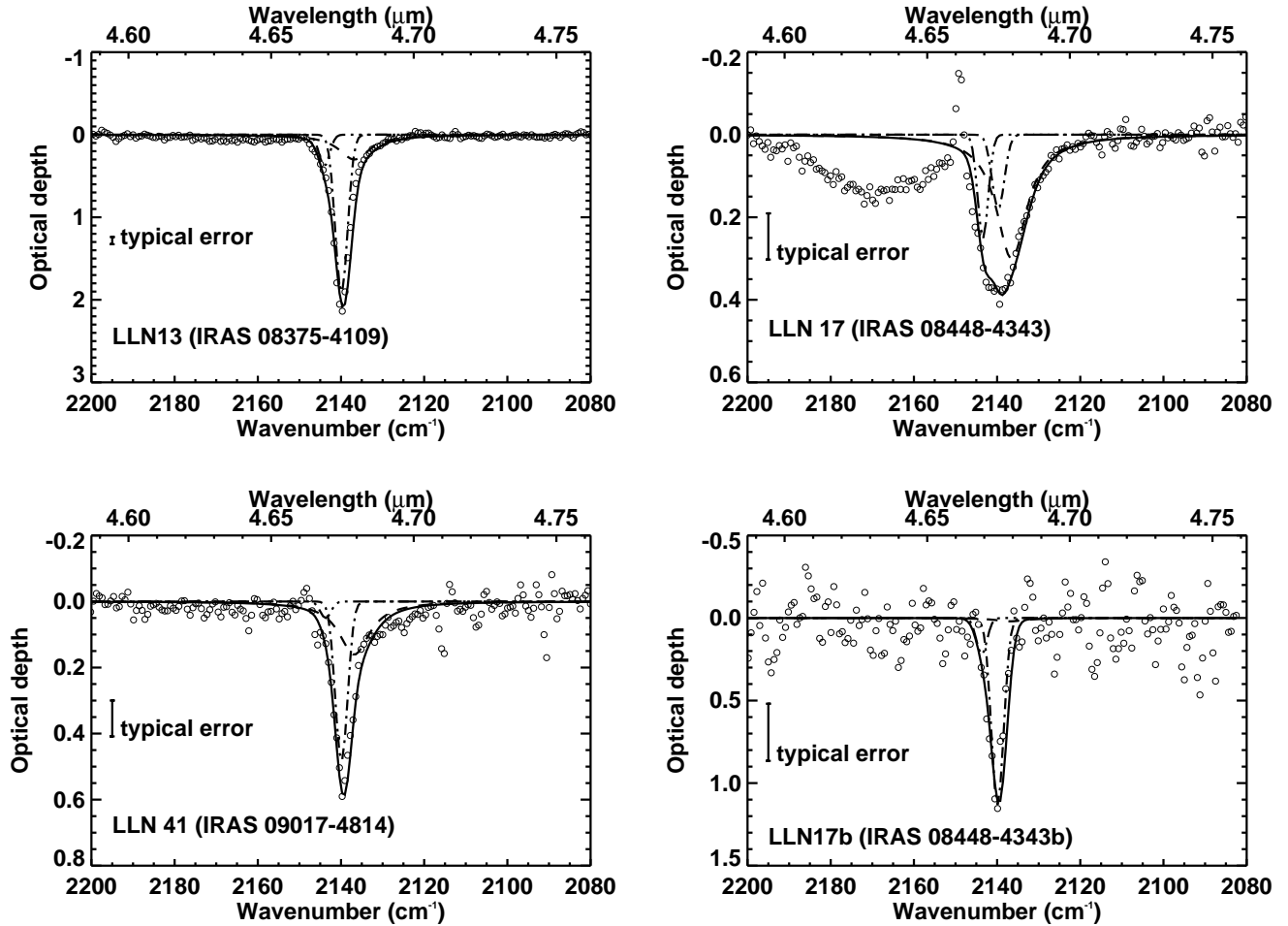}}
\caption{Best fit to the observed spectra using the three components phenomenological decomposition.
  The observed data points are plotted as open circles. The thick line
  is the sum of the three components after convolution by the profile of the ISAAC spectrometer ($R\simeq$~800).\label{vela:fig_co_components}.}
\end{figure*}

\begin{figure*}
\resizebox{\hsize}{!}{\includegraphics[angle=90]{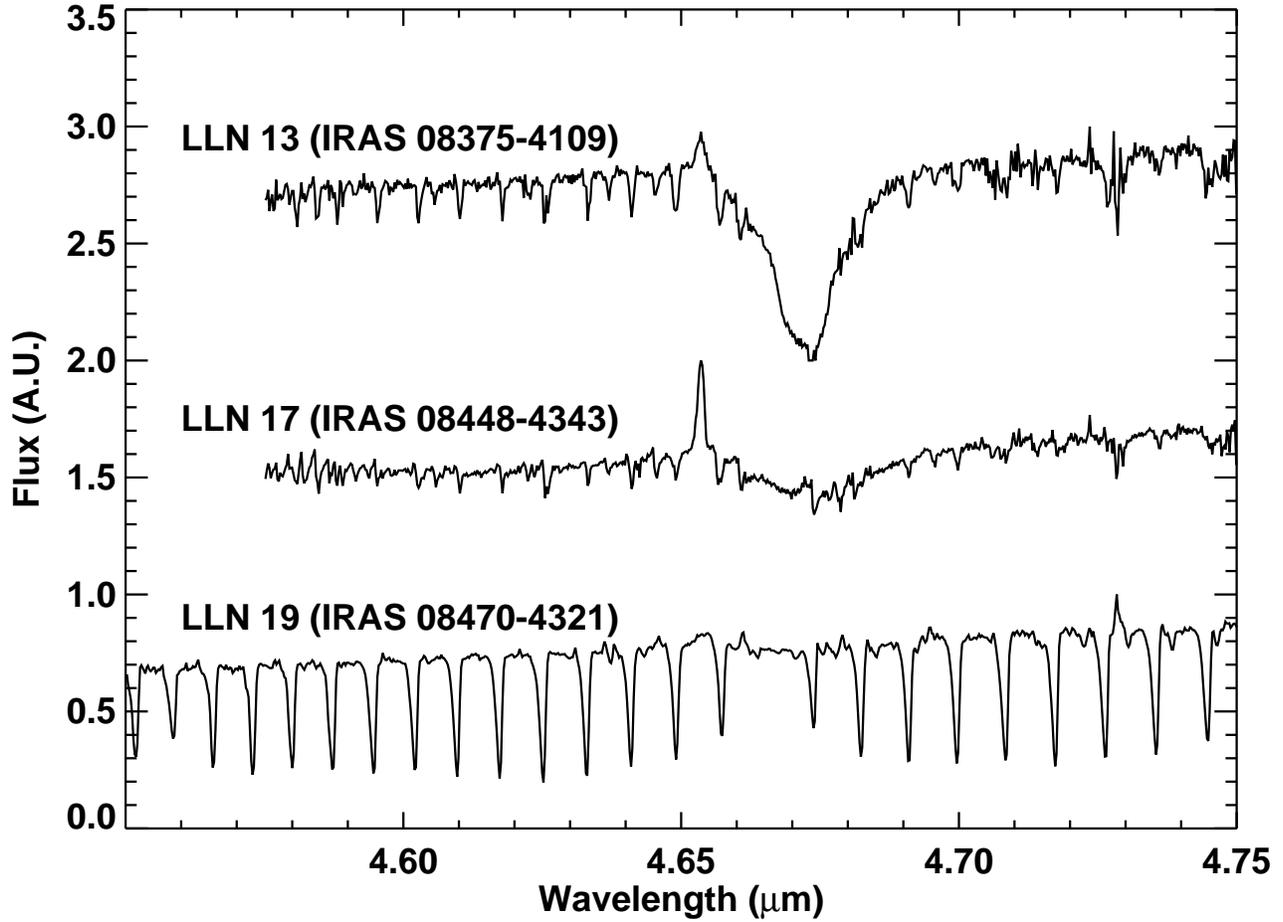}}
\caption{Medium resolution spectra ($R\simeq$~10,000) of three objects. The gas phase CO absorption lines are clearly visible and dominate the spectrum of \object{LLN 19}.\label{vela:fig_co_highres}}
\end{figure*}

\begin{figure}
\resizebox{\hsize}{!}{\includegraphics[angle=90]{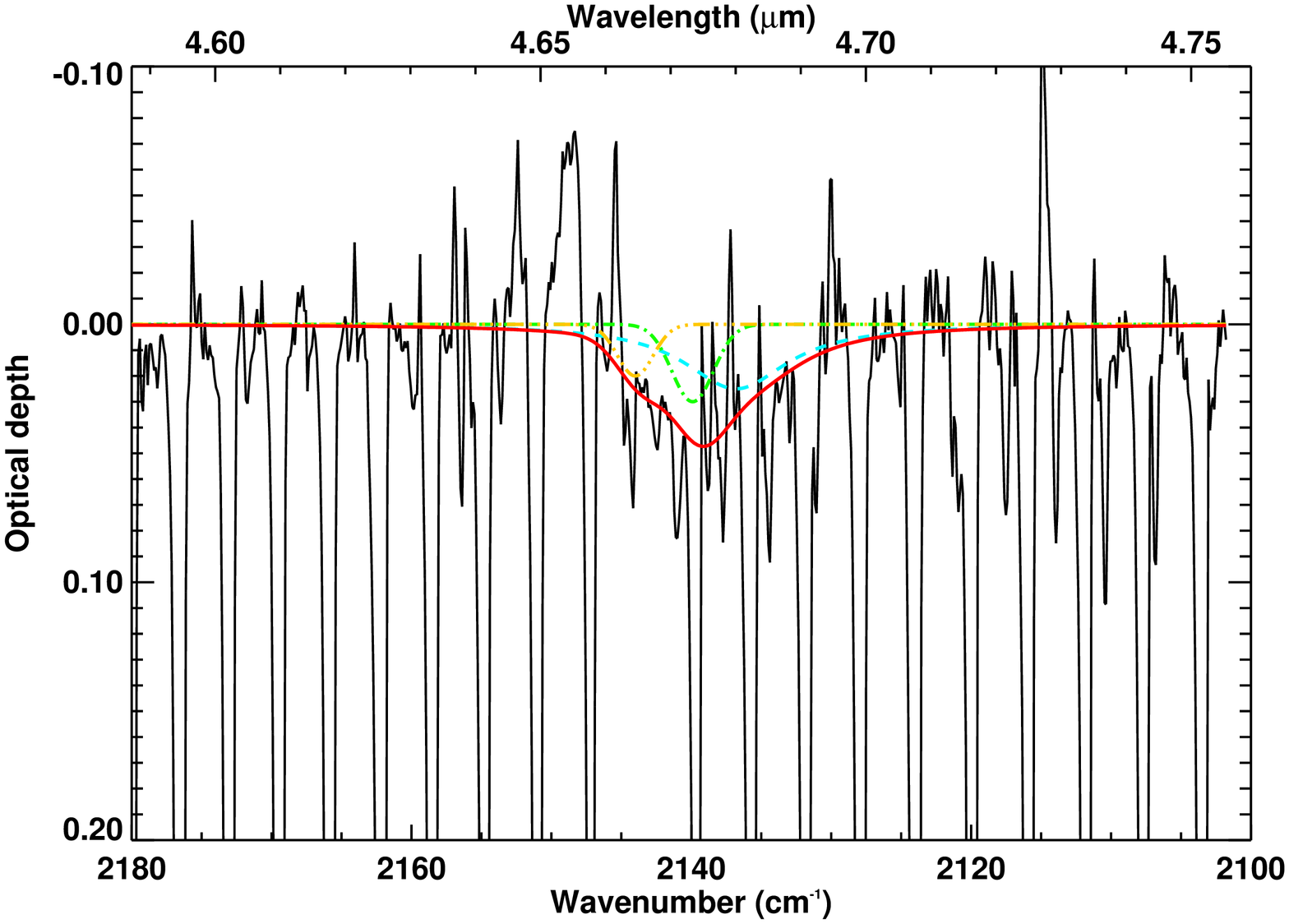}}
\caption{Continuum subtracted {\em VLT ISAAC} $M$-band spectrum of \object{LLN~19} around the CO ice feature at  2140 cm$^{-1}$ (4.675 $\mu$m). The spectrum is in an optical depth scale. The thick (red) line is the sum of the three fitting components.\label{vela:fig_co_lln19_highres}}
\end{figure}

In the $M$-band atmospheric window (4.5-4.8~$\mu$m), one can expect to
detect the solid CO absorption feature and/or the ro-vibrational band
of gaseous CO. A strong feature centered around 4.67~$\mu$m (2141
cm$^{-1}$) corresponding to CO-ice is detected in three out of the
five sources, namely \object{LLN 13} (\object{IRAS 08375--4109}),
\object{LLN 17} (\object{IRAS 08448--4343}) and \object{LLN 41} ({IRAS
  09017-4716}) and one of the companion objects \object{LLN 17b}
({IRAS 08448-4343b}). Their spectra are plotted in
Fig.~\ref{vela:fig_co_components} (open circles).  The gas-phase CO
ro-vibrational lines can introduce error in the interpretation of the
solid feature.  In particular, the P(1) line of gas-phase CO lies at
the center of the CO-ice band.  Medium resolution spectra obtained
toward three objects (\object{LLN 13}, \object{LLN 17} and
\object{LLN19}) are presented in Fig.~\ref{vela:fig_co_highres}. CO
ice is detected in the medium , but not in the low resolution spectrum
(Fig.~\ref{vela:fig_co_lln19_highres}) toward \object{LLN~19}.  The
cause of the discrepancy is the strong CO gas phase absorption lines,
which render the detection of the weak CO ice feature problematic.

In contrast to the water-ice profile, the FWHM of the solid-CO feature
differs from source to source (4.6--12.2 cm$^{-1}$). The peak
positions lie in a narrow range (2139--2141 cm$^{-1}$). The FWHM seen
toward \object{LLN 13} (\object{IRAS 08375--4109}), \object{LLN 41}
(\object{IRAS 09017-4716}) and \object{LLN 17b} (\object{IRAS
  08448--4343b}) are narrow (4.6--5.8 cm$^{-1}$). A narrow width is
often observed toward field stars and young embedded low-mass young
stellar objects (\citealt{Pontoppidan2003A&A...408..981P};
\citealt{Chiar1995ApJ...455..234C}). The source \object{IRAS
  08448--4343} (\object{LLN 17}) shows one of the broadest solid CO
features ever observed ($\Delta \nu$=12.2 cm$^{-1}$).  We fitted the
spectra using the decomposition outlined in
\cite{Pontoppidan2003A&A...408..981P} who showed that every
line-of-sight can be fitted by a linear combination of a narrow middle
component (mc), a broad red component (rc) and a blue component (bc).
The components are Lorentzian for the red component and Gaussian for
the blue and middle components. The central wavenumber and FWHM of
each component and their interpretation are given in
\cite{Pontoppidan2003A&A...408..981P}.  The phenomenological fits to
the spectra allow an astrophysical classification of the sources while
more classical fits with laboratory data help in understanding the
grain mantle composition as well as the grain shape and size source by
source. The best fits are found by a Simplex optimization method
(e.g., \citealt{Press1992nrfa.book.....P}). The measurement errors are
set constant with wavenumber so that an uniform weight is given to all
data points. 
  
The optical depths and derived columns densities are summarized in
Table~\ref{vela:co_components_tau}, while the abundances relative
  to H$_2$O ice are given in Table~\ref{Vela:table_ice_abundances}.
The solid CO column densities were obtained by integrating numerically
the three components. The band strength for pure CO ice was measured
to be $A$ = 1.1 $\times$ 10$^{-17}$ cm molec$^{-1}$ at 14~K
(\citealt{Jiang1975JChPh..62.1201J};
\citealt{Schmitt1989ApJ...340L..33S};
\citealt{Gerakines1995A&A...296..810G}). The uncertainties in the
measured $A$ value are not taken into account but are smaller than
that introduced by the continuum subtraction. Although the band
strength shows a 13\% variation when CO is mixed with other molecules
and a 17\% variation with increasing temperature
(\citealt{Schmitt1989ApJ...340L..33S};
\citealt{Gerakines1995A&A...296..810G}), the same value for $A$ is
adopted to compute the column density for the three components.  The
errors of $\sim$ 20--30\% reflect the noise in the observed spectra
and the telluric features removal.  Also presented in
Table~\ref{vela:co_components_tau} are the CO-ice upper limits for
\object{LLN 8} (\object{IRAS 08211-4158}). Most of the CO is located
in the gas for the latter object (see Fig.~\ref{vela:fig_co_highres}).
The analysis of the gas phase CO lines is presented in a separate
paper (Thi et al.\, in preparation). The column densities of the
middle and red component were estimated by integrating the each band
individually.  Again, a unique band strength of 1.1 $\times$ 10$^{17}$
cm molec$^{-1}$ is assumed.

\tiny
\begin{table*}
\centering
\caption[]{CO-ice band optical depth and column density derived from the phenomenological decomposition.
\label{vela:co_components_tau}}
\begin{tabular}[!ht]{lllllllll}
\noalign{\smallskip}
\hline
\hline
\noalign{\smallskip}
\multicolumn{1}{c}{Source}&\multicolumn{1}{c}{$N_{\mathrm{CO}}^{\mathrm{total}}$}& \multicolumn{1}{c}{$\tau^{\mathrm{bc}}_{\mathrm{CO}}$}&\multicolumn{1}{c}{$\tau^{\mathrm{mc}}_{\mathrm{CO}}$}&\multicolumn{1}{c}{$\tau^{\mathrm{rc}}_{\mathrm{CO}}$} & \multicolumn{1}{c}{$N_{\mathrm{CO}}^{\mathrm{bc}}$} & \multicolumn{1}{c}{$N_{\mathrm{CO}}^{\mathrm{mc}}$} &\multicolumn{1}{c}{$N_{\mathrm{CO}}^{\mathrm{rc}}$}& \multicolumn{1}{c}{$N_{\mathrm{CO}}^{\mathrm{mc}}$/}\\
                          &\multicolumn{1}{c}{(10$^{17}$ cm$^{-2}$)}&         &   &  & \multicolumn{1}{c}{(10$^{17}$ cm$^{-2}$)}&\multicolumn{1}{c}{(10$^{17}$ cm$^{-2}$)}  & \multicolumn{1}{c}{(10$^{17}$ cm$^{-2}$)} & \multicolumn{1}{c}{$N_{\mathrm{CO}}^{\mathrm{rc}}$}\\
\noalign{\smallskip}
\hline
\noalign{\smallskip}
\object{LLN 8}  & $<$\phantom{1}1.8                            &          $<$0.10           &    ...                       &         ...                  &      ...                  &      ...            &  ...           & ...\\
\object{LLN 13}   & \phantom{$<$}12.15$\pm$1.71  & \phantom{$<$}0.37$\pm$0.12 & \phantom{$<$}1.90$\pm$0.12 & \phantom{$<$}0.36$\pm$0.12 & \phantom{$<$}1.10$\pm$0.39 & \phantom{$<$}6.53$\pm$0.15  & \phantom{$<$}4.52$\pm$1.62& 1.44\\
\object{LLN 17}   & \phantom{$<$1}5.37$\pm$2.22  & \phantom{$<$}0.25$\pm$0.15 & \phantom{$<$}0.18$\pm$0.15 & \phantom{$<$}0.32$\pm$0.15 & \phantom{$<$}0.72$\pm$0.48 & \phantom{$<$}0.61$\pm$0.19  & \phantom{$<$}4.04$\pm$2.10& 0.15\\
\object{LLN 17b}  & \phantom{$<$1}3.83$\pm$1.74  &          $<$0.51           & \phantom{$<$}1.13$\pm$0.51 &          $<$0.51            &        $<$1.5           & \phantom{$<$}3.83$\pm$0.58  &         $<$6.39           & $>$0.6 \\
\object{LLN 19}$^b$   & $<$\phantom{1}1.8                            &          $<$0.10           &    ...                       &         ...                  &      ...                  &      ...            &  ...           & ...\\
\object{LLN 19}$^c$   & \phantom{$<$1}0.43$\pm$0.05     & \phantom{$<$}0.02$\pm$0.005          & \phantom{$<$}0.03$\pm$0.005         &  \phantom{$<$}0.025$\pm$0.005          &  \phantom{$<$}0.07$\pm$0.02            &  \phantom{$<$}0.10$\pm$0.02          & \phantom{$<$}0.25$\pm$0.05        & 0.41\\
\object{LLN 41}   & \phantom{$<$1}4.09$\pm$2.16  &          $<$0.15           & \phantom{$<$}0.48$\pm$0.15 & \phantom{$<$}0.18$\pm$0.15 & \phantom{$<$}0.14$\pm$0.15 & \phantom{$<$}1.64$\pm$0.18  & \phantom{$<$}2.27$\pm$2.04& 0.72\\
\object{LLN 41b}  & $<$\phantom{1}1.8                            &          $<$0.10           &    ...                       &         ...                  &      ...                  &      ...             &  ...           & ...\\
\object{LLN 20}$^d$ & \phantom{$<$1}5.19$\pm$0.96  & \phantom{$<$}0.12$\pm$0.07 & \phantom{$<$}0.38$\pm$0.08 & \phantom{$<$}0.29$\pm$0.04 &  \phantom{$<$}0.35$\pm$0.20     &  \phantom{$<$}1.29$\pm$0.27   & \phantom{$<$}3.55$\pm$0.49& 0.36\\
\object{LLN 33}$^d$ & \phantom{$<$1}7.85$\pm$0.81                 & \phantom{$<$}0.21$\pm$0.05 & \phantom{$<$}0.94$\pm$0.09 & \phantom{$<$}0.33$\pm$0.03 &  \phantom{$<$}0.61$\pm$0.14     &  \phantom{$<$}3.20$\pm$0.31  & \phantom{$<$}4.04$\pm$0.37& 0.79\\
\object{LLN 39}$^d$ & \phantom{$<$1}0.44$\pm$0.16                 & \phantom{$<$}0.004$\pm$0.003& \phantom{$<$}0.02$\pm$0.01 & \phantom{$<$}0.03$\pm$0.01 &  \phantom{$<$}0.011$\pm$0.008   &  \phantom{$<$}0.068$\pm$0.034 & \phantom{$<$}0.37$\pm$0.12& 0.18\\
\object{LLN 47}$^d$ &           $<$\phantom{1}0.36                &          $<$0.06           &         $<$0.02              &         $<$0.01           &    $<$0.17                        & $<$0.07                 & $<$0.12&  ... \\
\noalign{\smallskip}
\hline
\noalign{\smallskip}
\end{tabular}
\begin{flushleft}
$^a$The error bars are 3$\sigma$ level.\\
$^b$Low resolution spectrum.\\
$^c$Medium resolution spectrum.\\
$^d$Data from \cite{Pontoppidan2003A&A...408..981P}.\\
\end{flushleft}
\end{table*}
\normalsize

The line-of-sight toward \object{LLN 13} (\object{IRAS 08375-4109})
shows a high middle/red CO ratio typical of CO ice seen toward
low-mass protostars and field stars.  \object{LLN 41} (\object{IRAS
  09017-4716}) is also well matched by a high middle/red CO ratio
mixture. In contrast, \object{LLN 17} (\object{IRAS 08448-4343})
exhibits a stronger red component than a middle one. The CO-ice
profile and optical depth is similar to that found toward the
high-mass star \object{GL961E} \citep{Chiar1998ApJ...498..716C}
although the FWHM of $\sim$12.2 cm$^{-1}$ is among the largest ever
observed. The optical depths and column densities derived from the
fits with the phenomenological components are shown in
Table~\ref{vela:co_components_tau}, together with complementary data
obtained from \cite{Pontoppidan2003A&A...404L..17P}. The total CO ice
column densities using fits with laboratory data and phenomenological
components are similar within the errors, although the individual
components show larger variations.

\subsection{The 4.62~$\mu$m $''$XCN$''$ feature}
\label{vela:xcn}

Other weaker bands can also provide constraints on the energetic
processes in the vicinity of young protostars. In particular, the
so-called XCN band near 4.62 $\mu$m, contained in the CO survey, has
been cited as energetic processes tracer since its discovery toward
the massive protostar \object{W33A} \citep{Lacy1984ApJ...276..533L}.
The feature is particularly strong toward \object{LLN 17} (see
Fig.~\ref{vela:fig_co_components}), and is widely attributed to
vibrational stretching modes of -CN groups in molecular coatings on
dust grains.  The best candidate is the OCN$^{-}$ ion (e.g.,
\citealt{Novozamsky2001A&A...379..588N}), as first proposed by
\cite{Grim1987ApJ...321L..91G}. Quantitatively, adopting an integrated
band strength $A=$ (4--10) $\times$ 10$^{-17}$ cm molecule$^{-1}$ for
OCN$^{-}$ \citep{Demyk1998A&A...339..553D}, the feature corresponds to
a column density of 4.3-11 $\times$ 10$^{16}$ cm$^{-2}$.  Recent
laboratory work suggests however that $A$ may be closer to 1.3
$\times$ 10$^{16}$ cm molecule$^{-1}$
\citep{vanBroekhuizen2004A&A...415..425V}. The concentration relative
to H$_2$O using the latter value is $\sim$1\%. The derived column
density is similar to that found around other YSOs (e.g.,
\citealt{Whittet2001ApJ...550..793W};
\citealt{Demyk1998A&A...339..553D}). Upper limits are difficult to
estimate for the other objects because the wavelength range of the
OCN$^-$ feature is dominated by gas phase CO lines in emission or
absorption. The upper limits are given in
  Table~\ref{Vela:table_water_results} and the relative abundances in
  Table~\ref{Vela:table_ice_abundances}.  Noteworthy, OCN$^{-}$ is
detected in the object that shows the broadest CO feature, hence the
most processed,  but has one of the lowest CO/H$_2$O ratios in our
  sample. Further discussion on OCN$^{-}$ in \object{LLN 17} is
postponed to Sect.~\ref{peculiar_LLN17}.

\section{Discussion}
\label{vela:cloud}

\subsection{Companion objects}\label{companions_discussion}

Water ice is detected toward four companion objects. The column
densities are lower than those toward the primary object, except for
\object{LLN8b}.  The water ice may be located in the outer part of the
extended envelope surrounding both the main object and the companion.
CO ice has not been detected but the sample is too small and the upper
limit too high to allow further discussion, except for
\object{LLN~17}, which is further discussed in
Sect.~\ref{peculiar_LLN17}.

\subsection{Effect of pores on the water ice spectrum}\label{water_porous}

The ice layers on interstellar dust grains may be best simulated by
background deposition in the laboratory that forms a porous ice
(e.g., \citealt{Stevenson1999Sci}). A porous structure has several
advantages over the compact structure as a candidate for interstellar
ices.  A porous ice can retain a significant amount of molecules such
as CO. Moreover, by increasing the temperature, the adsorbed molecules
at the surface can migrate into the bulk where they are trapped, i.e.,
they remain in the mantle even if the grain temperature exceeds the
evaporation temperature of the species. The annealing increases the
diffusion of molecules into the pores. This point is further discussed
in Sect.~\ref{co_criterion}.

\begin{figure}[!ht]
  \resizebox{\hsize}{!}{\includegraphics[angle=90]{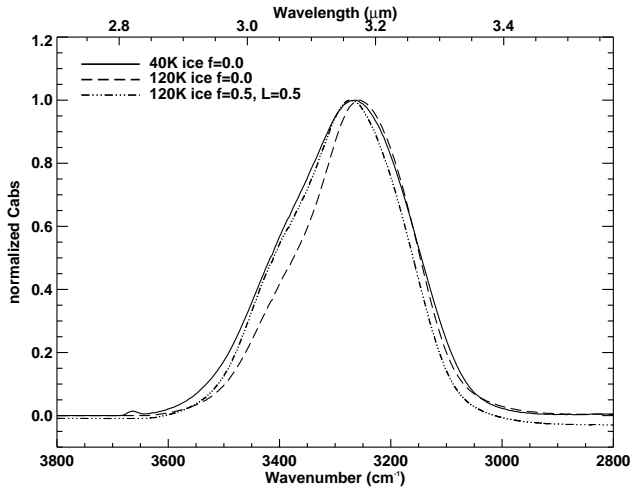}}
\caption{The effects of the depolarization factor $L$ on the water ice feature at  3300 cm$^{-1}$ ($\sim$~3.1~$\mu$m). The effect of porosity for water ice at 120~K is shown ($f$=0 means compact ice while $f$=0.5 means that 50\% of the volumn fraction is vacuum. The porous 120~K water ice absorption coefficient is close to the compact 40~K ice. \label{vela:fig_h2o7}}
\end{figure}

The effects on the optical constants of a water ice matrix due to
isolated (i.e. not connected with each other) inclusions or pores can
be simulated using an effective medium theory.  One of the
formulations is the Maxwell-Garnett approximation (e.g.,
\citealt{Bohren1983asls.book.....B} for a detailed description).  The
generalized Maxwell-Garnett formula with ellipsoidal pores is as
follows \citep{Niklasson1984JAP....55.3382N}:
\begin{equation}
\epsilon_{\rm eff}=\frac{f \epsilon _{\rm ice}(1- \epsilon _{\rm ice})(1-L)+\epsilon _{\rm ice}[L+(1-L)\epsilon _{\rm ice}]}{L+(1-L) \epsilon _{\rm ice}-f(1- \epsilon _{\rm ice})L}
\end{equation}
where $\epsilon_{\rm ice}$ is the complex dielectric function of the
ice, $f$ is the volume fraction of the inclusion (here the vacuum) and
$L$ is the depolarization factor. This factor is equal to $1/3$ for
spherical pores and lies between $1/3$ and 1 for prolate spheroidal
cavities (long axis along the normal to the layer and to the electric
field). By setting $L=1/3$, the classical Maxwell-Garnett formula is
recovered. The depolarization factor characterizes the effects of the
shape of the pores but does not model the differences in the chemical
bonding between a compact and a porous ice. The approximation was used
to study the influence of various porosities and the depolarization
factor on the simulated spectra.  The role of the depolarization
factor is studied by keeping all other parameters constant. The
resulting spectra are displayed in Fig.~\ref{vela:fig_h2o7}. The
simulations were performed with a porosity of 0.5 (i.e. 50\% of the
volume is vacuum). It is difficult to estimate the porosity of actual
interstellar water ice mantle, but 0.5 is most likely the highest
possible value. The generalized Maxwell-Garnett approximation is in
theory valid for porosities lower than $\sim$0.3. Simulations with
changing the porosity from 0.2 to 0.5 does not result, however, in
large differences in the shape of the profile. On the other hand, a
small variation in the depolarization factor (e.g.  from 0.3 to 0.5)
does. In Fig.~\ref{vela:fig_h2o7}, it can be seen that a higher
depolarization moves the peak of the absorption to shorter wavelengths
and increases the absorption at higher frequencies compared to that at
lower frequencies. This effect may jeopardize the estimate of the ice
temperature in Sect.~\ref{vela:L_band}.  The possible ambiguity
between a low-temperature and high-temperature ice with different
depolarization factors is illustrated in Fig.~\ref{vela:fig_h2o7}.
The simulated shapes are very similar in the blue.  The simulations
suggest that water ice between 10 and 40~K can fit the observed
profiles. However, the 120~K optical constant of porous ice is not able to mimic amorphous cold ice. Porous ice at 120~K lacks absorption in the red but
extra extinction can be provided by scattering by large grains.
It should be noted that the water ice feature at 3$\mu$m alone does not
provide sufficient constraints on the porosity of the ice.
Laboratory spectra of water ice at various porosity are warranted to
test the validity of the modeling.

It is known that some warm water ice can be hidden in the broad
profile \citep{Dartois2001A&A...365..144D}.  Crystalline H$_2$O ice
was observed in several lines of sight, mostly in the ejected envelope
of evolved stars
\citep{Smith1988ApJ...334..209S,Maldoni2003MNRAS.345..912M} and in the
circumstellar environment of the massive YSO BN object
\citep{Smith1989ApJ...344..413S}. It is known that the ice mantle on
grains around evolved stars is formed by condensation of water
molecules synthesized in the gas phase
\citep{Dijkstra2003A&A...401..599D}, whereas the water ice mantle of
interstellar grains is likely created by grain-surface reactions
\citep{Nguyen2002MNRAS.329..301N}. In conclusion, a significant amount
of high-temperature ($T_{\mathrm{ice}}$=40-60~K) ice can be hidden in
highly porous ice because of this effect, making the estimate of the
water ice temperature from the direct fit to the spectrum unreliable,
although there is no direct evidence of highly porous water ice in the
ISM. Further theoretical and laboratory investigations are needed to
constrain the degree of porosity in ice mantles.

\subsection{Evidence for thermal processing of water and CO ices}\label{co_criterion}

The sources chosen for this study were selected among the brightest
class I young stellar objects found in the Vela molecular clouds by
\cite{Liseau1992A&A...265..577L}. The objects posses similar
characteristics but exhibit a large variations in water and CO ice
column densities. It is therefore interesting to find possible
relations between the ice properties and the source characteristics.
The source characteristics of interest are the extinction $A_V$, the
infrared luminosity $L_{\mathrm{IR}}$, and colors such as $K-L$, IRAS
12~$\mu$m/25~$\mu$m flux ratio $R$(12/25) and 25~$\mu$m/60~$\mu$m flux
ratio $R$(25/60). The ratio $R$(12/25) is a good estimate of the warm
dust temperature (100~$<$~$T$~$<$~250~K), while the ratio $R$(25/60)
is sensitive to cooler dust (50~$<$~$T$~$<$~100~K). Finally, the $K-L$
color allows to probe hot dust ($T>$~250). The source characteristics
were chosen because they are available for all the sources in our
sample.

\subsubsection{Water ice}\label{water_ice_processing}

\begin{figure}[!ht]
\resizebox{\hsize}{!}{\includegraphics[angle=90]{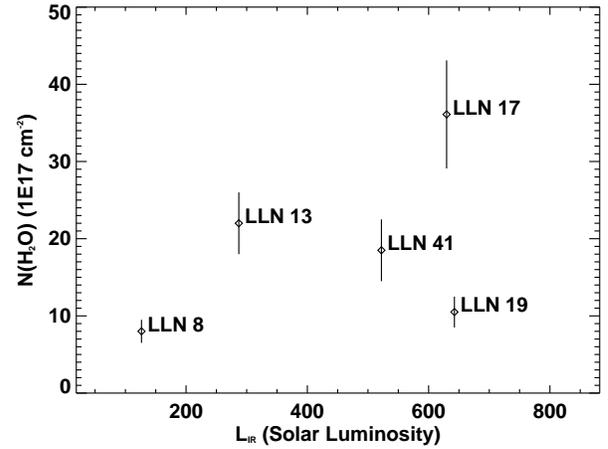}}
\caption{Water ice column density $N$(H$_2$O) versus the infrared luminosity $L_{\mathrm{IR}}$. The error bars are 2$\sigma$ level.\label{vela:fig_NH2O_LIR_Vela}}
\end{figure}

The water ice column densities are plotted against the infrared
luminosity in the upper left panel of
Fig.~\ref{vela:fig_NH2O_LIR_Vela}.
They vary from 0.8 to 3.6 $\times$ 10$^{18}$ cm$^{-2}$ and do not
  correlate with increasing infrared luminosity measured between 7 and
  135 $\mu$m.
\begin{figure}[!ht]
\resizebox{\hsize}{!}{\includegraphics[angle=90]{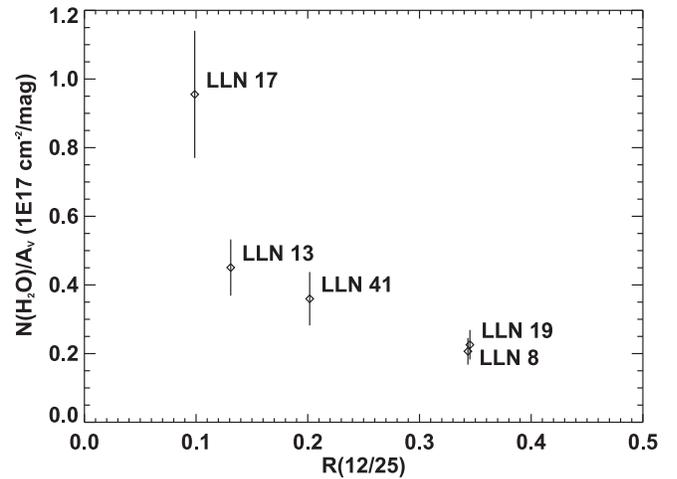}}
\caption{Water ice column density $N$(H$_2$O) normalized to the extinction A$_{\mathrm{V}}$ versus the ratio $R$(12/25). The error bars are 2$\sigma$ level. The errors on the ratios $R$(12/25) are $\sim$ 5--10 \%. A trend is seen between $N$(H$_2$O)/A$_{\mathrm{V}}$ and $R$(12/25). \label{vela:fig_NH2O_r12_25_Vela}}
\end{figure}

\begin{figure*}[!ht]
\resizebox{\hsize}{!}{\includegraphics[]{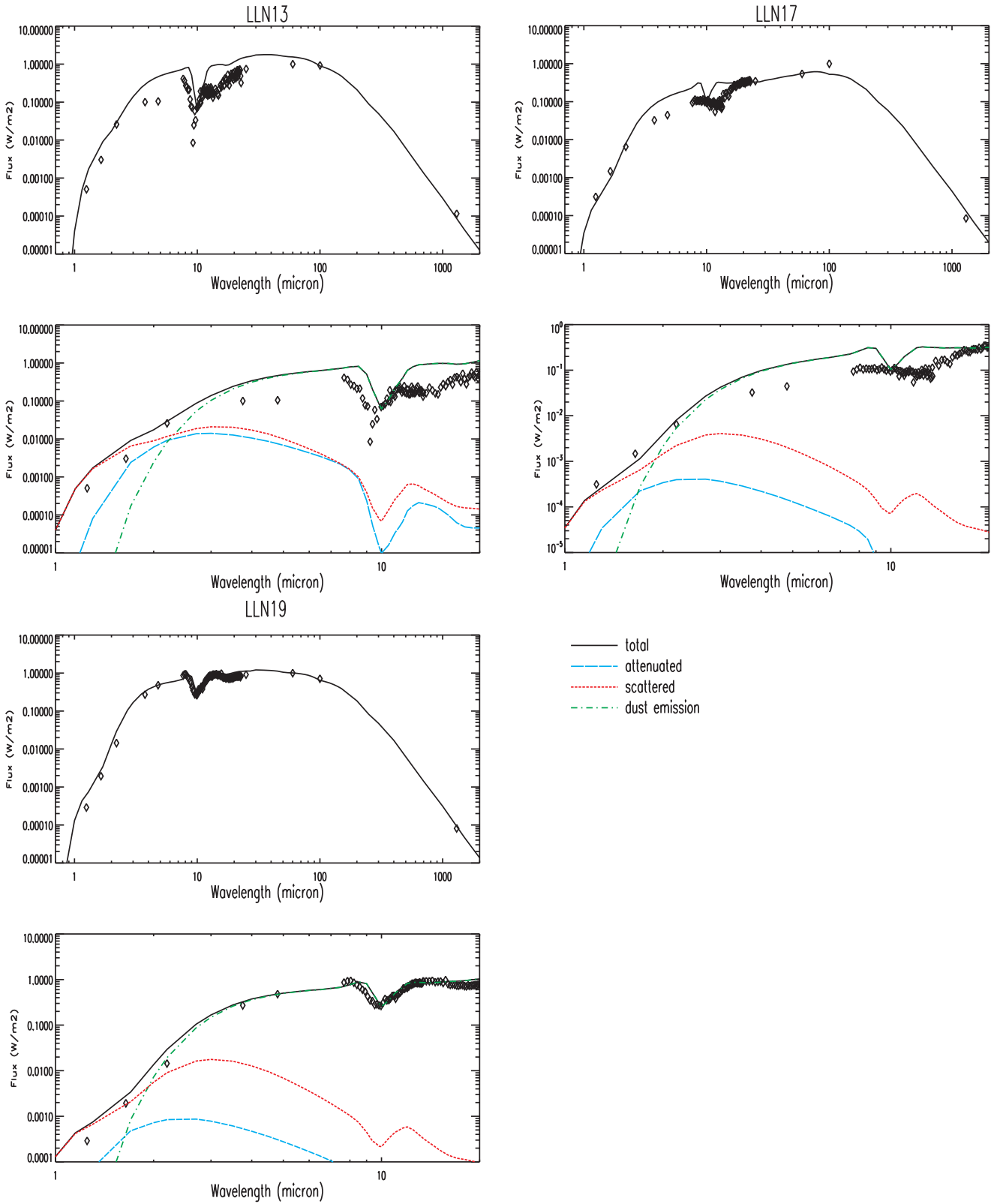}}
\caption{ Fit to the SED of \object{LLN 13}, \object{LLN 17} and \object{LLN 19}. The upper panels show the fit to the entire SED while the upper panels focus on the near to mid-infrared range. Also shown are the contribution by the extincted central object, the dust scattered light and the dust thermal emission.  \label{fig_dusty_sed}}
\end{figure*}

  The extinction can be estimated from the observed $E(H-K)$ with
  $A_{\mathrm{V}}$= (15.3 $\pm$ 0.6) $\times$ $E(H-K)$
  \citep{Rieke1985ApJ...288..618R}. The intrinsic $H-K$ color of early
  B up to F stars is close to zero. Assuming that the $K$ band flux is
  dominated by the extincted central source, $E(H-K)$ $\simeq$ $H-K$.
  The estimated extinction is {\em an upper limit} since the $H-K$
  color may be lower if emission from hot circumstellar dust at
  $\sim$~1500~K dominates the near-infrared continuum. Alternatively,
  the optical depth at 9.7 $\mu$m (A$_{\mathrm{V}}$=(18.5 $\pm$ 1.5)
  $\times$ $\tau_{9.7}$, \citet{Roche1985MNRAS.215..425R}), measured
  in the IRAS-LRS spectra, provides a lower limit on the extinction
  because of potential intrinsic silicate emission. Therefore, the
  combined use of the two methods brackets the actual extinction. The
  different estimates of the extinction $A_{\mathrm{V}}$ are provided
  in Table~\ref{table_Av}. \cite{Liseau1992A&A...265..577L} estimated
  lower limits of 20 magnitudes for most sources in our sample. The
  discrepancy between the two methods is about a factor 3. In order to
  better estimate $A_{\mathrm{V}}$, the full Spectral Energy
  Distribution has been modelled together with the IRAS-LRS data between 9 and 25
  $\mu$m of three objects (\object{LLN~13}, \object{LLN~17} and
  \object{LLN~19}) using the one-dimensional public
  radiative transfer code DUSTY \citep{Ivezic1997MNRAS.287..799I}
  adopting a bare silicate grain model. The three objects were chosen
  because they have measured 1.3 mm fluxes. The best fits to the SEDs
  are shown in Fig.\ \ref{fig_dusty_sed} and the estimated
  $A_{\mathrm{V}}$ are given in Table~\ref{table_Av}.  DUSTY fails to
  fit the details of the silicate feature of (\object{LLN~13},
  \object{LLN~17}), but provides reasonably good fit of the silicate
  feature for \object{LLN~19}. The absorption in 10--25 $\mu$m region
  of \object{LLN~13} and \object{LLN~17} can be caused by water and
  CO$_2$ ice in addition to silicate absorption.  The near-infrared
  and the SED fitting method give relatively close values for
  $A_{\mathrm{V}}$. Thus, we will adopt the simpler
  near-infrared method in the rest of the discussion. The $H$- and
$K$-band magnitudes are included in Table~\ref{vela:table_objects}.
The derived extinctions using $E(H-K)$ are between 20 and 50,
consistent with the values derived from the silicate feature.

\begin{table}
\centering
\caption[]{Estimation of the visual extinction by the 9.7 $\mu$m optical depth, the near-infrared excess and the SED fitting method.\label{table_Av}}
\begin{tabular}[!ht]{llll}
\noalign{\smallskip}
\hline
\hline
\noalign{\smallskip}
                          & \multicolumn{3}{c}{Method}\\
\cline{2-4}\\
\multicolumn{1}{c}{Source}& \multicolumn{1}{c}{"$\tau_{9.7}$"} & \multicolumn{1}{c}{"$E(H-K)$"} & \multicolumn{1}{c}{"SED fitting"}\\
\noalign{\smallskip}
\hline
\noalign{\smallskip}
\object{LLN 8}  & $>$~8   & $<$~39 & ...\\
\object{LLN 13} & $>$~48  & $<$~59 & 60\\
\object{LLN 17} & $>$~13  & $<$~46 & 40 \\
\object{LLN 19} & $>$~22  & $<$~56 & 45 \\
\object{LLN 41} & $>$~10  & $<$~40 & ...\\
\noalign{\smallskip}
\hline
\noalign{\smallskip}
\end{tabular}
\end{table}

Adopting the relation $N_{\mathrm{H}}$=(1.6 $\times$ 10$^{21}$) $\times$
$A_{\mathrm{V}}$ cm$^{-2}$, the derived water ice abundance is
(1.3--5.9) $\times$ 10$^{-5}$ and is lower than the mean value found
in quiescent molecular clouds in Taurus (7 $\times$ 10$^{-5}$,
\citealt{Whittet2003dge..conf.....W}) and Serpens (9 $\pm$ 1 $\times$
10$^{-5}$, \citealt{Pontoppidan2004A&ASerpens}). The lower water ice
abundances found toward the YSO's in Vela may be ascribed to thermal
processing of the dust grains by the central object, lower
  average gas density, or by external radiation from surrounding
YSO's.  \cite{Jorgensen2004A&A...416..603J} show that variations in
the gas-phase abundances in the envelopes of YSOs can be explained by
a variable size of the region over which the molecules are frozen out;
a similar situation may apply here.

To further ascertain the effect of thermal heating, the water ice
abundance $N$(H$_2$O)/$A_{\mathrm{V}}$ is plotted against the IRAS
12~$\mu$m/25~$\mu$m flux ratio $R$(12/25) in Fig.
\ref{vela:fig_NH2O_r12_25_Vela}. The IRAS 12~$\mu$m band filter is
relatively narrow and therefore, the IRAS fluxes are not strongly
affected by the silicate absorption band.  The water ice abundance
clearly decreases with increasing warm dust temperature. This is
  consistent with the work of \citet{Boogert2000A&A...353..349B}, who
  were the first to show that the color temperature of the dust
  correlates with ice heating/crystallization through the CO$_2$ ice
  bands, which trace CO$_2$ ice mixed with H$_2$O ice. The heating of
the water ice mantle at $T>$100~K should also results in the formation
of a significant amount of heated water ice (narrower profile).
However, the water ice observed in our sources is mostly in an
amorphous state (see Sect.~\ref{vela:L_band}), although the narrow
feature of warm water ice may be broadened in porous ices
(see Sect.~\ref{water_porous}). Alternatively, the inspection of the IRAS-LRS spectra
  toward \object{LLN~13} and \object{LLN~17} shows that the 12$\mu$m
  photometry can be biased by water ice absorption. Indeed, the
  presence of large amounts of water ice with respect to silicate will
  decrease the ratio $R$(12/25) and explain the trend seen in
  Fig.~\ref{vela:fig_NH2O_r12_25_Vela}.

\subsubsection{CO ice}

\begin{figure*}[ht]
  \centering
  \resizebox{\hsize}{!}{\includegraphics[angle=90]{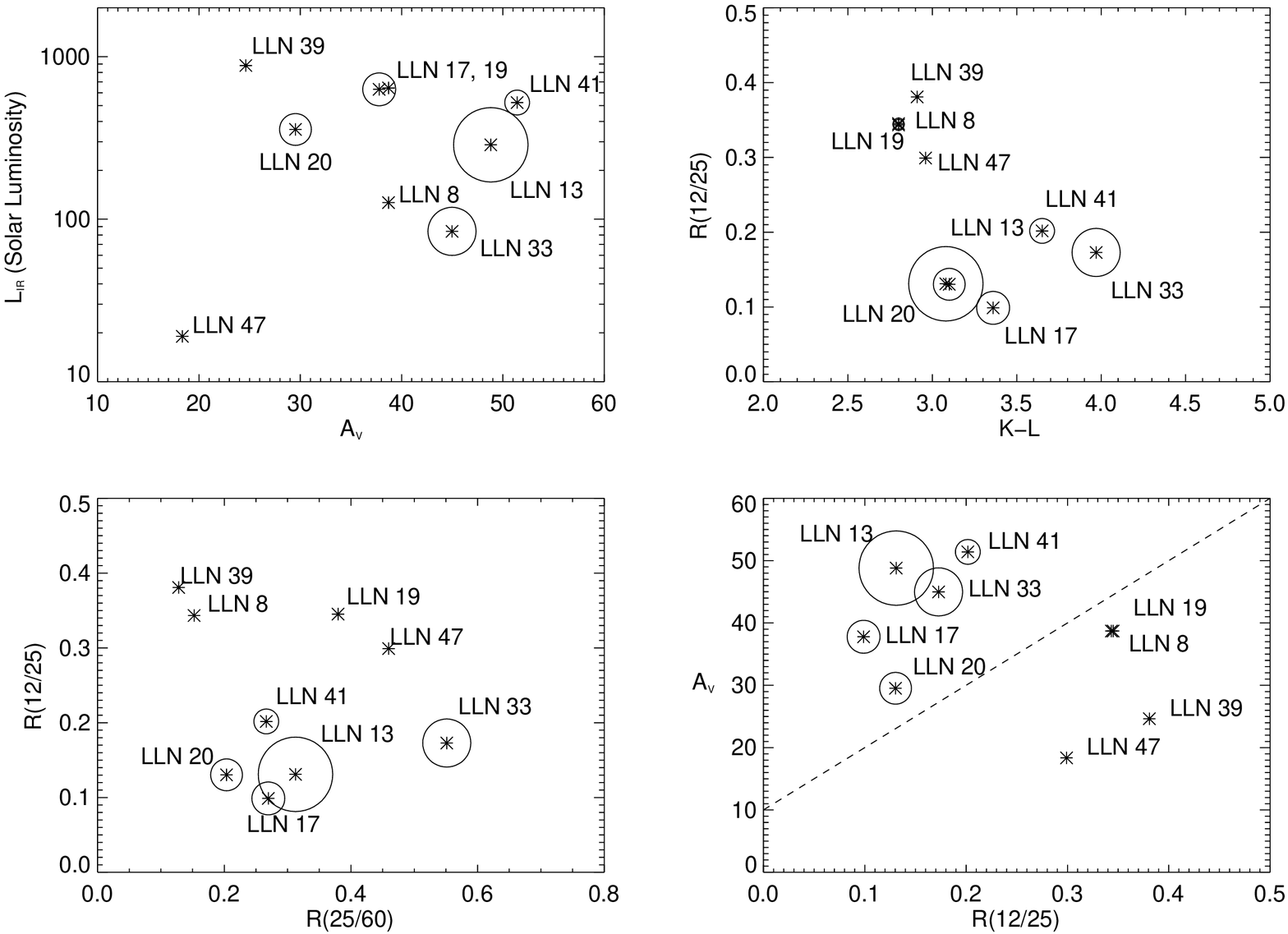}}
\caption{
  The upper left panel shows the infrared luminosity of the sources
  computed using the IRAS 4 bands fluxes versus an estimate of
  the extinction toward the objects. The radius of the circles at the
  position of each object is proportional to the observed abundance of
  CO ice. The upper right and lower left panels are two color-color
  diagrams. The infrared ratios $R$(12/25) and $R$(25/60) are ratios
  between the fluxes $F$(Jy) at 12 and 25 $\mu$m and between fluxes at
  25 and 60 $\mu$m respectively. The lower right panel is a extinction
  versus $R$(12/25) diagram.  The dashed-line represents a possible
  diving line between presence and absence of CO ice.
  \label{vela:fig_color_diag}}
\end{figure*}

In this subsection, we added four objects located in the Vela cloud
studied by \cite{Pontoppidan2003A&A...408..981P}. A closer look at the
source characteristics and their possible relationships with CO ice
abundance are given in Fig.~\ref{vela:fig_color_diag}.  In this
figure, the abundance of CO ice for each object is represented by a
circle whose radius is proportional to the CO ice column density.  The
upper left panel of Fig.~\ref{vela:fig_color_diag} shows the
line-of-sight extinction $A_{\mathrm{V}}$ and infrared luminosity
$L_{\mathrm{IR}}$ of the nine sources.

The total column density of CO ice does not correlate with the
infrared nor the bolometric luminosity of the sources. In particular
two sources \object{LLN 17} and \object{LLN 19} show similar values
for $A_{\mathrm{V}}$ and $L_{\mathrm{IR}}$ but CO ice has only been
detected toward \object{LLN 17}. 

It may be possible to constrain the dust temperature range to which
the CO ice abundance is sensitive by finding relations between CO ice
abundances and colors. The $K-L$, $R$(12/25), and $R$(25/60) colors
encompass dust with temperature ranging from 50 to more than 250~K.
This range of dust temperature is well above the CO ice sublimation
temperature, thus we do not expect any correlation apart, from
perhaps, a weak one with the coolest temperature $R$(25/60).  From the
two color-color diagrams plotted in Fig.~\ref{vela:fig_color_diag}
(the upper-right and lower-left panels), it appears that the CO ice
abundances vary with the ratio $R$(12/25) but not with $K-L$, which is
expected, nor with $R$(25/60), which is more surprising. Before
discussing this finding, it is important to test whether the CO ice
abundance is related to the extinction in the line-of-sight. Water ice
column densities are known to vary linearly with the visual extinction
after a certain threshold value (e.g.,
\citealt{Whittet2003dge..conf.....W}). In other words, water ice
abundances do not change with $A_V$. For the CO ice in our sample, the
abundances do not correlate with $A_V$ as testified by the lower-right
panel of Fig.~\ref{vela:fig_color_diag}.
  
\begin{figure}[!ht]
\resizebox{\hsize}{!}{\includegraphics[angle=90]{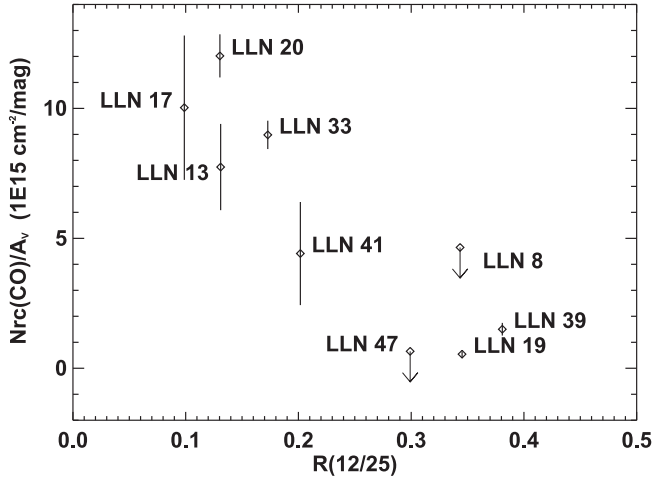}}
\caption{Red component column density $N_{\mathrm{rc}}$ normalized to the extinction A$_{\mathrm{V}}$ versus the ratio $R$(12/25). The error bars are 2$\sigma$ level. The errors on the ratios $R$(12/25) are $\sim$ 5--10 \%. A clear trend is seen between $N_{\mathrm{rc}}$/A$_{\mathrm{V}}$ and $R$(12/25). \label{vela:fig_Nrc_r12_25_Vela}}
\end{figure}

Pure CO ice sublimes at $\sim$20~K, while CO trapped in water ice,
often associated with the red component at 2136 cm$^{-1}$, evaporates
with sudden phase changes of the water matrix at higher temperature
\citep{Schmitt1989pmcm.rept...65S,Collings2003ApJ...583.1058C}.  The
amount of CO ice trapped with water should therefore show a similarly
decreasing trend as water ice with increasing value of $R$(12/25). The
column density of the red component normalized to the estimated
extinction $N_{\mathrm{rc}}$(CO)/$A_{\mathrm{V}}$ is plotted in
Fig.~\ref{vela:fig_Nrc_r12_25_Vela}.
$N_{\mathrm{rc}}$(CO)/$A_{\mathrm{V}}$ decreases with increasing value
of $R$(12/25) . Both $N_{\mathrm{rc}}$(CO)/$A_{\mathrm{V}}$ and
$N$(H$_2$O)/$A_{\mathrm{V}}$ show the same trend with $R$(12/25) and
therefore correlate with each other (right panel of
Fig.~\ref{vela:fig_Nrc_water}). This correlation is consistent with
the simultaneous sublimation of water and CO molecules trapped in the
water matrix. It is also clear that $N$(CO)/$A_{\mathrm{V}}$ and
$N$(H$_2$O)/$A_{\mathrm{V}}$ do not correlate (left panel of
Fig.~\ref{vela:fig_Nrc_water}). If $R$(12/25) is another way to
  express $N$(H$_2$O)/$A_{\mathrm{V}}$, then
  Fig.~\ref{vela:fig_Nrc_r12_25_Vela} and
  Fig.~\ref{vela:fig_Nrc_water} express the same relationship between
  the CO red ice component and water ice.
  
It is difficult to solely attribute the red component to trapped CO
molecules because of the absence in space of the 2152 cm$^{-1}$
feature, which is present in laboratory data of CO-H$_2$O mixtures.
However, the 2152 cm$^{-1}$ absorption feature seen in laboratory
spectra of CO-water mixtures disappears when the dust is heated above
$\sim$80~K \citep{Fraser2004MNRAS.353...59F}. Other effects may
  contribute to the red component. One possibility is that part of
the red component is due to scattering by the larger grains in the
grain size distribution similar to the effect seen for the red wing in
the water ice band (see Sect.~\ref{vela:L_band} and
\citealt{Dartois2005A&A}).  In this scenario, the CO and water ice may
located in two layers, which ensures that the 2136 cm$^{-1}$ feature
does not appear even at low temperature. The CO ice trapped in a water
matrix and the pure CO ice in large grains may be located in two
separate populations in the line-of-sight. The CO red component
appears to be a good temperature indicator. The possible relationship
between $N_{\mathrm{rc}}$(CO)/$A_{\mathrm{V}}$ and $R$(12/25) should
be tested in other star-forming regions. Likewise, future observations
of $^{13}$CO$_2$ ice will allow to test the possible relation between
heated $^{13}$CO$_2$ ice and the fraction of hot over cold grains for
intermediate-mass YSOs. {\em Spitzer} observations of CO$_2$ ice
  toward the low-mass protostar \object{HH~46} have shown that most of the
  ices are located in the cold ($T$~$<$~50~K) part of a circumstellar
  envelope, although the inner envelope should be warm enough to
  explain the presence of a double-peaked absorption feature for the
  CO$_2$ bending mode \citep{Boogert2004ApJS..154..359B}. The best fit
  to the CO$_2$ bending mode feature is obtained with an ice mixture
  CH$_3$OH:H$_2$O:CO$_2$=0.3:1:1 at 155~K in the laboratory or
  $\sim$~75~K in space. This indicates that the inner H$_2$O, CO$_2$ and
  CH$_3$OH ices are warm.

\subsection{Is \object{LLN~17} a peculiar intermediate mass YSO?}
\label{peculiar_LLN17}

Among the intermediate-mass YSOs observed in the Vela Molecular Cloud,
\object{LLN~17} is the only object where large quantities of
H$_2$O-rich CO ice, of methanol ice and of OCN$^{-}$ are found;
  although it should be emphasized that the relative abundance of
  methanol ice toward \object{LLN~17} is not extremely high
  ($\simeq$~6.9~\% w.r.t. H$_2$O ice). The ice abundances seen
toward \object{LLN~17} may be attributed to its peculiar environment.
\object{LLN~17} is the most luminous YSO in our sample, it is located
in the densest stellar cluster (see Table~\ref{vela:table_objects})
and shows signs of molecular outflows
\citep{Lorenzetti2002ApJ...564..839L,Giannini2005A&A...433..941G}.  In
this section, we discuss whether these particularities can explain the
ice abundances.

First we discuss the different ways to form solid-CH$_{\mathrm{3}}$OH
and their implications.  The chemistry of solid-CH$_{\mathrm{3}}$OH is
still subject to considerable discussion. The exact moment of its
formation and the amount of subsequent energetic processing are
currently unknown.  Originally it was suggested that methanol ice
formed through hydrogenation of accreted CO
\citep{Tielens1987ip...symp..397T}, but recent laboratory experiments
suggest that this mechanism has a very low yield (e.g.
\citealt{Hiraoka2002ApJ...577..265H};
\citealt{Hiraoka2005ApJ...620..542H}), although the results are not
conclusive (e.g., \citealt{Watanabe2004ApJ...616..638W};
\citealt{Watanabe2002ApJ...571L.173W}). Hydrogen addition to CO at low
temperature most likely occurs by H-tunneling since this reaction has
an energetic barrier \citep{Woon2002ApJ...569..541W}.  The original
scheme allows an efficient synthesis of methanol ice in cold
pre-stellar cores where CO molecule is known to be highly depleted
onto grains although the amount of atomic H, which reacts with CO to
form CH$_3$OH, is also small \citep{vanderTak2000A&A...361..327V}. If
methanol ice is formed at an earlier phase of star formation, probably
at the pre-stellar core stage without significant energetic processing
afterwards apart from the processing by UV photons generated by
cosmic-rays interacting with H$_2$ and H, similar abundances should be
found for members of the same cluster, which is not the case for
objects in the Vela molecular cloud and in the Serpens cloud
\citep{Pontoppidan2003A&A...404L..17P}. Likewise, methanol ice is not
present in most lines of sight in the Taurus cloud
\citep{Chiar1996ApJ...472..665C}. Therefore, methanol ice is probably
synthesized at a later stage of stellar collapse where energetic
events (UV and/or warm-up in shocks created from the interaction of
stellar wind/outflow with the surrounding envelope) from the central
object can provide the necessary energy to overcome the reaction
barrier.  \object{LLN~17} is the most massive YSO in our sample and,
hence, the most UV luminous.  It would be tempting to attribute the
CH$_3$OH ice to grain surface reactions triggered by UV photons from
the central object. However, if we take into account all YSOs where
methanol ice has been found, it is also clear that the methanol ice
abundance does not correlate solely with the UV flux from the central
star and direct stellar UV processing is also not dominant (e.g.,
\citealt{Dartois1999A&A...342L..32D};
\citealt{Pontoppidan2003A&A...404L..17P}). In addition to the UV from
the central star, shocks can produce copious amount of UV photons and
heat the gas and dust. \cite{Bergin1998ApJ...499..777B} have shown
that large amounts of water ice can form in post-shock regions.
According to this scenario, the water ice mantle is formed by
condensation of water molecules formed in the high temperature region
of shocks, which subsequently condense rapidly onto the bare silicate
grains in the cool post-shock region. In theory, the water ice formed
by vapor deposition at T~$>$~100~K should be in the crystalline form.
As discussed previously, crystalline water ice is rarely seen toward
YSOs. One possibility to circumvent the problem is that high water 
deposition rate can result in amorphous instead of crystalline ice
  \citep{Kouchi1994A&A...290.1009K}.  Another possibility is that
crystalline ice is amorphized by cosmic-ray bombardment or Ly $\alpha$
radiation \citep{Leto2003A&A...397....7L}.  As realized by
\citet{Bergin1998ApJ...499..777B,Bergin1999ApJ...510L.145B}, current
gas phase chemical networks do not include all possible high
temperature formation paths of methanol. As a consequence, the amount
of methanol formed in shocks in standard models is too low to account
for the abundance in the solid phase. Another possibility to form
efficiently CH$_3$OH ice is that hot atomic hydrogens created by
dissociation of H$_2$ in shocks impinge onto the grain surface to
react with CO to form HCO, H$_2$CO and finally CH$_3$OH. Chemical
models of shocks that include all formation paths of molecules at high
temperature are warranted.

The methanol ice found in \object{LLN~17} is most likely embedded into
a water matrix (Sect.~\ref{vela:methanol}), indicating that water and
methanol ice may have formed simultaneously or that methanol molecules
have migrated. Interestingly, a collimated bipolar H$_2$ jet structure
composed of bright knots of line emission has been detected toward
\object{LLN~17} while neither \object{LLN 19} nor \object{LLN~13} show
sign of outflow \citep{Lorenzetti2002ApJ...564..839L}.  The morphology
of the \object{LLN~17} jet points to the presence of episodic
phenomena of mass ejection typically observed in protostellar jets
\citep{Reipurth1992A&A...257..693R}. The molecular outflow has been
mapped by \cite{Wouterloot1999A&AS..140..177W} in the $^{12}$CO 1-0
transition. Important unknowns in the outflow/shock scenario are the
degree of mixing of the processed and unprocessed grains, the
precession and the periodicity of the jet phenomenon. In summary,
detailed models are needed to determine whether shock chemistry is a
viable mechanism to synthesize large amounts of gas and solid phase
CH$_3$OH. Likewise, correlation between outflow and high methanol ice
content should be further investigated.

\begin{figure}[!ht]
\resizebox{\hsize}{!}{\includegraphics[angle=90]{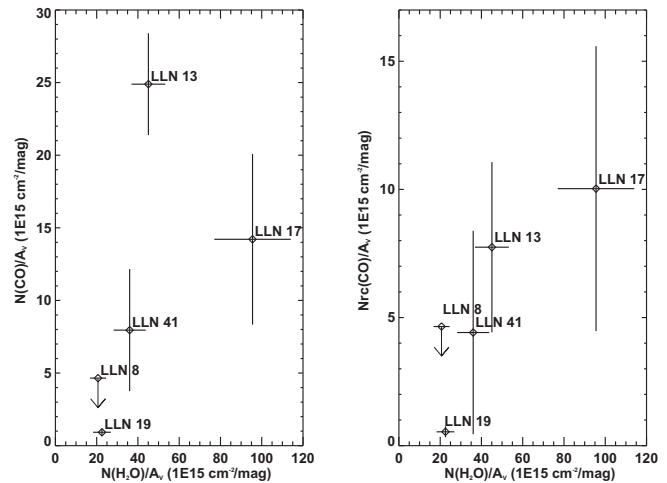}}
\caption{Total CO column density $N$(CO) and red component column density $N_{\mathrm{rc}}$ normalized to the extinction A$_{\mathrm{V}}$ versus $N$(H$_2$O)/A$_{\mathrm{V}}$ in the left and right panel respectively. A clear trend is seen between $N$(CO)$_{\mathrm{rc}}$/A$_{\mathrm{V}}$
and $N$(H$_2$O)/A$_{\mathrm{V}}$. \label{vela:fig_Nrc_water}}
\end{figure}

It is also interesting to relate the stellar and circumstellar
characteristics of \object{LLN 17} to the presence of OCN$^{-}$ in its
envelope. The OCN$^{-}$ ion is believed to be formed in a solid state
acid-base reaction between NH$_3$ and isocyanic acid (HNCO),
efficiently mediated by thermal processing
\citep{Demyk1998A&A...339..553D,Novozamsky2001A&A...379..588N,
  Park2004ApJ...601L..63P,vanBroekhuizen2004A&A...415..425V}.  The
main difficulty in this interpretation is that isocyanic acid ice has
never been detected in space, although it has a relatively high
gas-phase abundance in hot cores and in shocked regions
\citep{Comito2005ApJS..156..127C}.  The high abundance in shocked
regions is consistent with the proposed outflow/shock scenario.

Theoretically, HNCO ice abundance up to $\sim$3\% with respect to
water ice can also be attained by grain-surface reactions
\citep{Hasegawa1993MNRAS.263..589H}.  Alternatively, an abundance of
$\sim$1\% for OCN$^{-}$ can be easily reached by UV-photolysis of the
initial mixture H$_2$O/CH$_3$OH/NH$_3$=100:15:15
\citep{vanBroekhuizen2004A&A...415..425V}.  Although the required
amount of UV-dose (fluence) exceeds the value estimated in dense
clouds, the large amount of processed CO and CH$_3$OH ice found toward
\object{LLN~17} suggests that the UV field around \object{LLN~17} may
be enhanced compared to that provided by cosmic rays induced only. The
abundance of NH$_3$ in the envelope around \object{LLN~17} is
difficult to estimate owing to the poor signal-to-noise ratio of the
spectrum around the ammonia feature at 2.97 $\mu$m. However,
sufficient amounts of NH$_3$ may have existed in the ice mantle since
the 3.47 $\mu$m feature is relatively strong toward \object{LLN~17},
assuming that the 3.47 $\mu$m feature is caused by ammonia hydrate.
Noteworthy, the detection of OCN$^{-}$ is often concomittent with that
of CH$_3$OH ice (see Table 5 of
\citet{vanBroekhuizen2004A&A...415..425V} and the references therein).

Another particularity of \object{LLN~17} is that the water ice
abundance with respect to H$_2$ is the highest (see
Fig.~\ref{vela:fig_Nrc_water}) in our sample. Methanol ice and OCN$^-$
are often detected in objects with high water ice abundance
\citep{vanBroekhuizen2004A&A...415..425V,Pontoppidan2003A&A...404L..17P,Dartois1999A&A...342L..32D}.

Finally, the presence of water-rich CO ice in \object{LLN~17} and its
absence in the companion object \object{LLN~17b} as seen in Figure
\ref{vela:fig_co_components} reinforces the idea that the processing
occurs in the vicinity of the central object (i.e. with a few thousand
AU). The $M$-band spectrum of \object{LLN~17b} is too low to give
meaningful upper limit on the amount of OCN$^{-}$. The difference
between \object{LLN 17} and its companion suggests that external
heating and/or UV processing by the other stars in the cluster play a
minor role (see last column of Table~\ref{vela:co_components_tau}).

 Interestingly, the OCN$^-$ and methanol abundances relative to
  water in \object{LLN~17} are close to those found in the envelope
  around \object{HH 46} \citep{Boogert2004ApJS..154..359B}. Likewise,
  the water-rich CO abundance dominates over the water-poor CO and water ice is
  relatively abundant (5.7 $\times$ 10$^{-5}$).

In summary, the simultaneous presence of large amount of water-rich CO ice,
CH$_3$OH ice , and OCN$^{-}$ ice toward \object{LLN~17} and the
non-detection in other objects are consistent with the idea that these
species are formed by a combination of UV and thermal processing in
the inner regions of the circumstellar envelope. The UV radiation can
be generated by the interaction between the outflows and envelope, by
the central source, or by a combination of both mechanisms. Another
possibility is that CH$_3$OH is first synthesized in the gas-phase,
then condenses onto grains simultaneously with H$_2$O in post-shocked
regions.

\section{Conclusions}
\label{vela:conclusion}

We have obtained $L$- and $M$- band spectra of a sample of
intermediate-mass protostars in the Vela molecular cloud with the
VLT-ISAAC. This is the first significant sample of intermediate mass
protostars for which ice data are published.

A broad absorption feature at $\sim$3.01 $\mu$m is detected in all
sources (main and companion objects). The features show an extended
wing beyond 3.25 $\mu$m, which can be reproduced in part by
  scattering by grains at radius 0.4--0.5 $\mu$m. The water ice
feature is dominated by absorption from cold amorphous ice although
the spectroscopic signature of warm water ice can be masked if the ice
is porous.

Methanol ice is only detected around the protostar \object{LLN 17}
(\object{IRAS 08448--4343}). The derived abundance is 10 $\pm$ 2~\%
relative to water ice. The upper limit on the methanol abundance
toward the other sources is between 5 and 10\% with respect to water
ice.

Solid CO is detected in four main objects and one companion object.
The profiles show a large variety of shapes. A strong variation of the
total CO ice column density is found. We decompose the CO ice feature
into three components. The column of CO becomes significant (i.e.
larger than 10$^{17}$ cm$^{-2}$) only at $R$(12/25) greater than
$\sim$0.3. The color $R$(12/25) may trace the abundance of water
  ice with respect to silicate. There is no clear trend between the
column density of pure CO ice, traced by the middle component, and
other characteristics of the YSOs ($L_{\mathrm{bol}}$,
A$_{\mathrm{V}}$, IRAS bands flux ratios). On the other hand, we find
a possible correlation between the ratio of the flux at 12~$\mu$m and
25~$\mu$m, $R$(12/25), which is a measure of the warm dust temperature
(100$<$ $T$ $<$ 250~K), and the amount of CO ice trapped in a water
rich ice mantle, traced by the red component.  Likewise, the amount of
CO ice trapped in a water and that of water may correlate.  This
possible correlation is consistent with the idea that the water ice
and the CO embedded in it sublime simultaneously.

These possible correlations should be tested in other star-forming
regions. However, CO ice trapped in water-rich ice cannot solely
account for the large amount of CO in the red component seen toward
YSOs in Vela. Other factors such as scattering by the larger grains in
the size distribution can probably contribute to the red component.

A strong absorption feature centered at 4.62 $\mu$m is detected toward
\object{LLN 17} (\object{IRAS 08448--4343}). The feature is likely
caused by OCN$^-$.  The derived abundance relative to water is
1~$\pm$~0.2~\%. This feature is not detected in any other object in
our sample. Together with the detection of methanol and the broad CO
feature, the detection of OCN$^{-}$ suggests that the ice has been
thermally and/or UV processed in \object{LLN 17}. The processing
generated by UV from the central object is not essential and perhaps
shocks induced processing is at play. Further theoretical
investigations on the possibility to form large amounts of methanol
ice together with water ice in post-shocked regions are needed.
  
  Observations of a larger sample of high signal-to-noise ratio
  spectra obtained with 8-10 meter class telescopes (VLT, Gemini,
  Keck) and with Spitzer at longer wavelengths of protostars of
  varying luminosities combined with sophisticated laboratory
  experiments will improve our understanding of the nature of the ices
  and their role in the synthesis of complex molecules in the
  interstellar medium.

\begin{acknowledgements} \label{vela:acknowledgements}
  This work was supported by the Netherlands Organization for
  Scientific Research (NWO) grants 614.041.003 and 614.041.005 and a
  Spinoza grant. WFT was supported by a PPARC grant during his stay at
  UCL. WFT acknowledges an ESA internal fellowship. We thank the ESO
  staff for their help before and during the observations, in
  particular, F. Comeron, C.\ Lidman and O.\ Marco.  The analysis of
  the data has benefited from discussions with Annemieke Boonman, Doug
  Johnstone, Guillermo Mu\~noz-Caro, Xander Tielens and Chris Wright.
  Stimulating advice was provided by the late Prof.\ J.M.\ Greenberg.
  Finally, we thank the referee for her/his detailed comments which
  help to improve the paper.
\end{acknowledgements}


\bibliographystyle{aa}
\bibliography{vela} 

\end{document}